\documentclass[showkeys,floatfix,noshowpacs, preprintnumbers, aps,superscriptaddress,pra,10pt,twocolumn,longbibliography,nofootinbib]{revtex4-2}

\usepackage[utf8]{inputenc}

\usepackage{graphicx}
\usepackage{tikz}
\usetikzlibrary{shadows.blur}

\usepackage{physics}
\usepackage{braket}
\usepackage{float}  
\usepackage[english]{babel}
\usepackage[utf8]{inputenc}

\usepackage[colorlinks=true,linktocpage=true]{hyperref}
\hypersetup{
    allcolors  = {blue},
}
\usepackage{booktabs} 

\usepackage{amsmath}
\usepackage{amssymb} 
\usepackage{bbold}
\usepackage{amsthm}

\usepackage{xfrac}
\usepackage{mathtools}

\usepackage{standalone}

\usepackage{thmtools}
\usepackage{thm-restate}

\usepackage{listings}
\usepackage{enumitem}   

\usepackage[page,toc,titletoc,title]{appendix}

\usepackage{footmisc}
\usepackage{subcaption}

\usepackage{microtype} 
\usepackage[capitalize]{cleveref} 

\usepackage{nomencl}
\makenomenclature

\usepackage{soul}
\newtheorem{theorem}{Theorem}

\usepackage{csquotes}

\usepackage{etoolbox}

\patchcmd{\thebibliography}
  {\url}
  {\nolinkurl}
  {}{}

\renewcommand{\nolinkurl}[1]{}

\usepackage{xcolor}
 \usepackage{todonotes}

\usepackage{soul}

\begin{document}

\title{Wigner's friend's black hole adventure: an argument for complementarity?}
\author{Laurens Walleghem}\thanks{\newline Email: \href{laurens.walleghem@york.ac.uk}{laurens.walleghem@york.ac.uk}}
\affiliation{Department of Mathematics, University of York, Heslington, York YO10 5DD, United Kingdom}
\affiliation{International Iberian Nanotechnology Laboratory (INL), Av. Mestre Jos\'{e} Veiga, 4715-330 Braga, Portugal}\date{\today}

\begin{abstract}
At the heart of both black hole physics and Wigner’s friend scenarios lies the question of unitarity. 
In Wigner’s friend setups, sealed-lab measurements are modeled unitarily, probing the measurement problem.
In black hole physics, the unitarity problem concerns information preservation in evaporation. 
We extend a recent analogy between these two puzzles exposed by Hausmann and Renner [arXiv:2504.03835] by constructing new paradoxes that merge black hole physics with extensions of the Wigner’s friend scenario into a unified argument. 
This unified construction allows us to sharpen the black hole cloning paradox, which leaves room for a theory to consistently describe the physics of black holes satisfying the assumptions of this paradox but without operational contradictions concerning quantum theory's predictions.
We close this loophole, showing that no such theory exists, conditional on standard assumptions of black hole puzzles.
We moreover show how the firewall argument can be used to derive a violation of no-signalling.
We conclude by briefly highlighting important subtleties in assumptions commonly used in black hole puzzles, questioning the validity of these assumptions.
\end{abstract}

\date{\today}

\maketitle

\section{Introduction} \label{sec:introduction}

Arguably, few phenomena in the universe are as captivating as black holes, where two of the most foundational theories of modern science -- general relativity and quantum theory -- intersect in profound ways. 
Applying principles of quantum (field) theory to black holes, Hawking proved in his seminal paper~\cite{hawking1975particle} that black holes radiate, and by energy conservation thereby evaporate over time. 
However, Hawking’s calculations suggested this radiation to be approximately thermal, which seemed to indicate a loss of information: an initial pure state of matter collapsing to a black hole would evolve nonunitarily to a mixed state of radiation after evaporation~\cite{hawking1976breakdown}.
This enigma is also referred to as the black hole information  puzzle~\cite{harlow2016jerusalem,Polchinski_2016}.  

Page gave a general argument of how a black hole system might evolve, based on the assumption that the black hole and radiation combined are always in a random pure state \cite{page1993average,page1993information}. 
The resulting curve for the (renormalised~\cite{holzhey_geometric_1994,harlow2016jerusalem}) entanglement entropy\footnote{We distinguish in general two types of entropy: the entanglement entropy, also referred to as the von Neumann or fine-grained entropy, describing the entanglement of one system with another, and the thermodynamic entropy, a coarse-graining thereof~\cite{almheiri2021entropy}.} is estimated to decrease somewhat past halfway of the evaporation~\cite{page1993information,page2013time}, where it diverges from the entanglement entropy per Hawking's calculation.

From Page's unitary entropy curve, late-time radiation should be highly entangled with the early radiation. Moreover, the equivalence principle and quantum field theory (QFT) in curved spacetime suggest high entanglement across the horizon~\cite{braunstein2013better,reeh1961bemerkungen,eisert2010colloquium,Witten_2018,Hollands_2018,harlow2016jerusalem,mathur2009information}. Therefore, late-time radiation just outside the black hole should be highly entangled with both the early radiation and modes inside the black hole, apparently violating monogamy of entanglement, also known as the firewall paradox~\cite{braunstein2007quantum,braunstein2013better,almheiri2013black,mathur2009information}.

Another black hole puzzle is a cloning argument analysed by Susskind and Thorlacius~\cite{susskind1994gedanken} and later by Hayden and Preskill~\cite{hayden2007black}, suggesting a form of black hole complementarity~\cite{Stephens_1994,susskind1993stretched,susskind1994gedanken}. 
The firewall argument is sometimes presented as an argument against complementarity~\cite{harlow2016jerusalem}, but additionally assumes the approximate validity of semiclassical gravity to argue for high entanglement across the horizon.

In the field of quantum foundations, recent extensions~\cite{brukner2017quantum,frauchiger2018quantum,bong2020strong,cavalcanti2021implications,haddara2022possibilistic,utreras2023allowing,haddara2024local,szangolies2020quantum,walleghem_refined_2024,walleghem2023extended,walleghem2024connecting,walleghem2025extendedwignersfriendnogo,vilasini2019multi,vilasini2022general,ormrod2023theories,leegwater2022greenberger,schmid2023review,brukner2018no,ormrod2022no,nurgalieva2018inadequacy,montanhano2023contextuality,zukowski2021physics,walleghem2023extended,montanhano2023contextuality,guerin2020no,ying2023relating} of Wigner's friend scenario~\cite{wigner1995remarks}, by combining it with Bell nonlocality~\cite{brunner_bell_2014,bell1964einstein,Bell_1976} and contextuality~\cite{budroni_kochen-specker_2022,kochen1990onthe,spekkens2005contextuality,abramsky2011sheaf,schmid2020structure}, have caused much excitement, providing deep implications for the nature of reality. 
As with the black hole information puzzle, Wigner's thought experiment~\cite{wigner1995remarks} also concerns a question of unitarity, in this case probing the measurement problem by modeling an observer measuring a system unitarily~\cite{Maudlin1995,schlosshauer2005decoherence,legget20005measurement,Hance_2022}.
Extensions thereof combine the outcomes and perspectives of different observers to arrive at contradictions that force us to rethink combinations of seemingly innocent assumptions.

Inspired by recent work of Hausmann and Renner~\cite{hausmann2025firewallparadoxwignersfriend}, who draw analogies between Wigner's friend and black hole paradoxes, we combine extensions of the Wigner's friend scenario with black hole physics to derive no-go theorems that strengthen the black hole cloning paradox.
More specifically, the cloning paradox leaves room for a theory that could describe the entire black hole physics of the interior and exterior while satisfying the assumptions of this paradox but without (operational) contradictions concerning quantum theory's predictions. 
We close this loophole, showing that no such theory exists.
We also give an argument of how the firewall paradox assumptions give rise to a violation of no-signalling. 
We will assume standard (but often vaguely stated) assumptions of the cloning and firewall paradoxes for our protocols regarding unitarity and the decoding of Hawking radiation, but provide a more critical assessment in \Cref{sec:implications}. 
We will highlight key issues in these assumptions that might invalidate these black hole puzzles.
The logical role of these theorems is to show that the standard assumptions cannot be rescued by weakening quantum theory in these ways; hence one of the black hole assumptions must fail, and \Cref{sec:implications} identifies the candidates.
The focus in this paper lies on 4-dimensional asymptotically flat Schwarzschild black holes, but our arguments can be extended to other types of black holes.

In the remainder of this work, we recap Wigner's thought experiment in \Cref{sec:WF} and present our main protocol in \Cref{sec:combining_FR_and_BH}. 
We relate to literature in \Cref{sec:literature}, and conclude by highlighting some subtleties underlying black hole puzzles.

\vspace{0.5cm}

\section{Short recap of Wigner's friend} \label{sec:WF}

In Wigner's thought experiment, a friend measures a quantum system inside a sealed lab. If one assumes some universality of quantum theory, the friend must ultimately be a quantum system as well. 
Therefore, Wigner, a \textit{superobserver} outside the sealed lab, models the friend's measurement as a unitary interaction of quantum systems.
Specifically, suppose the friend measures a qubit system $S$, initialized in some state $\ket{\varphi_0} = \alpha \ket{0}+\beta\ket{1}$. 
The friend measures the system $S$ in the computational basis\footnote{By the computational or $Z$-basis of a qubit, we mean the $\ket{0},\ket{1}$ basis, and by $X$-basis the $\ket{\pm}$-basis.}. 
Let the state of the friend's lab, including the friend's memory, be denoted by $\ket{0}_F, \ket{1}_F$ when the friend obtains the outcome $f=0$ or $f=1$, respectively. Assuming the friend's lab is initialized in the $\ket{0}_F$ state,\footnote{Another initial state orthogonal to $\ket{0}_F,\ket{1}_F$ works as well~\cite{walleghem_refined_2024}, but for simplicity we take this to be the initial state.}
the friend's measurement can be modeled unitarily as a CNOT with control on $S$, denoted by $U_F$.
Namely, if the friend finds the system $S$ in the $\ket{0}$ state, she keeps her memory record at $f=0$, whereas if she finds the system $S$ in the $\ket{1}$ state, she changes her memory to $f=1$: \begin{equation}
    \ket{\varphi_0}_S \ket{0}_F  \overset{U_F}{\mapsto} \alpha \ket{00}_{SF}+\beta \ket{11}_{SF}.
\end{equation}

An actual human playing the role of the friend may be practically infeasible, but an artificially intelligent program on a future quantum computer may be a reasonable physical realisation of such an observer~\cite{wiseman2022thoughtful}.\footnote{Abstractly, we can think of an observer as an information processing system, potentially with a preferred basis to interpret its knowledge~\cite{walleghem_refined_2024,brukner2021qubits}.}
In our protocols and no-go theorems, we assume that the superobserver can perform arbitrary quantum operations on an observer and its environment, using the wording of Ref.~\cite{bong2020strong}.
The assumption that such superobservers exist can be argued to be not much more exotic than the capability of an observer to reconstruct the initial state of a complicated quantum system from its Hawking radiation.

Among the most celebrated extended Wigner's friend (EWF) arguments are the Frauchiger--Renner (FR) paradox~\cite{frauchiger2018quantum}\footnote{An implementation of the Frauchiger--Renner paradox~\cite{frauchiger2018quantum} in which the crucial quantum degrees of freedom would be gravitational is problematic, particularly with respect to diffeomorphism invariance and gravitational shielding, as discussed in Ref.~\cite{Dukehart:2024bsn}, but could, in principle, still be implemented in other degrees of freedom, electromagnetically or in a quantum computer, for example.} and Local Friendliness (LF) no-go theorem~\cite{bong2020strong,cavalcanti2021implications,haddara2022possibilistic} (see Appendix A for a short recap).
In the FR paradox, observers model each other quantumly and reason about each other's outcomes, arriving at a contradiction. 
We will use a refined version thereof~\cite{walleghem_refined_2024} that leads to a stronger no-go theorem.  
The LF no-go theorem is a more device-independent version of the FR paradox, and derives a contradiction between absolute outcomes and a notion of locality applied to outcomes of performed measurements~\cite{cavalcanti2021implications,bong2020strong}.
We next apply alterations of these extended Wigner's friend scenarios to black hole physics.

\vspace{0.5cm}

\begin{figure*}[t]
         \centering
\includestandalone[width=0.8\textwidth]{figures/setup_and_penrose}
         \caption{Initial schematic set-up of our first protocol (a) and pictured in a Penrose diagram (b) for a black hole formed from collapse. Alice and Bob measure $S_A,S_B$ and fall into the black hole. Ursula and Wigner catch their Hawking radiation (yellow wiggly arrow) upon which they perform an operation, after which they fall into the black hole to retrieve the outcomes of Alice and Bob (green wiggly arrow). As pointed out in \Cref{sec:critical_note_assumptions} and discussed further in \Cref{sec:implications}, one may question, among others, whether the location of the retrieval and decoding of Hawking radiation can be well into the bulk, whether Ursula and Wigner can still fall into the black hole after this decoding, and the correct causal structure. We also provide a theorem (\Cref{th:nogo_EWFcloning_notmeet}) where Ursula and Wigner remain outside the black hole.}
         \label{fig:radiation_mirror_timelike_path}
\end{figure*}

\section{Combining extended Wigner's friend paradoxes and black hole cloning}
\label{sec:combining_FR_and_BH}

In this section, we construct a protocol that combines the Frauchiger--Renner and black hole cloning paradox and derive two no-go theorems. 
We briefly compare to the cloning paradox. 
Implications 
are discussed in \Cref{sec:implications}.
Before we start, we comment upon assumptions from standard black hole puzzles that we adopt, and point out critical subtleties therein that we will discuss later in \Cref{sec:implications}.

\subsection{Critical note about black hole assumptions} \label{sec:critical_note_assumptions}

In the remainder of this section and in \Cref{sec:firewall_recap}, we provisionally adopt the standard assumptions used in black-hole cloning and firewall arguments. 
These include the existence of suitable black hole radiation subsystems, the possibility of decoding them in the required spacetime region, an approximate inside/outside factorization, persistent interior records, and the identification of late radiation with near-horizon modes. These assumptions are not derived here and are known to be conceptually delicate in quantum gravity. The no-go results below should therefore be read as conditional results about the strengthened consequences of these assumptions. In \Cref{sec:implications} we return to these assumptions and highlight key issues that cast doubt upon the validity of these assumptions. 

A finite-dimensional quantum-information model of the operationally relevant subsystems is assumed, with qubits and finite memories representing effective or code-subspace degrees of freedom. 
The no-go results are therefore conditional on the possibility of identifying, decoding, and coherently manipulating such effective subsystems, including subsystems encoded in the Hawking radiation.
From the viewpoint of an observer outside the black hole, the black hole dynamics is modeled, for the purposes of the protocol, as implementing a unitary scrambling/encoding map $U_R$ on the black hole, infalling systems and relevant environmental degrees of freedom, à la Hayden–Preskill~\cite{hayden2007black}. 
It is further assumed that an appropriate recovery operation on the accessible Hawking radiation (possibly together with previously collected radiation or auxiliary systems) reconstructs the relevant code-subspace state of the laboratory, after which the inverse of the laboratory measurement can be applied as in the Wigner's friend scenario. 
Where (and when) and how such information recovery would occur is not prescribed, and these questions constitute an important point of scrutiny, as noted above and further discussed in \Cref{sec:implications}. 

Finally, we assume that Alice and Bob hold systems that are sufficiently isolated within their lab and can be effectively modeled as qubits, upon which they will perform measurements. 
(This last assumption concerns only the relevant effective degrees of freedom, and does not require their gravitational fields to be perfectly shielded from their surroundings, for instance.)


\subsection{Protocol and no-go theorem} \label{sec:FR_BH_protocol}

Alice and Bob, residing in sealed labs, each measure their part of a Hardy state\footnote{We will sometimes omit the tensor product symbol for ease of notation, i.e. write $\ket{00}$ or $\ket{0}\ket{0}$ for $\ket{0} \otimes \ket{0}$.}~\cite{hardy1993nonlocality} \begin{equation} \label{eq:hardy} \ket{\psi_0}_{S_A S_B} = \frac{1}{\sqrt{3}}(\ket{00}+\ket{10}+\ket{11})_{S_A S_B} \end{equation} in the computational basis with outcomes $a,b \in \set{0,1}$, modeled unitarily as $U_A,U_B$, and next fall into an old black hole, i.e. a black hole past the Page time. Ursula and Wigner retrieve and decode the Hawking radiation in which the state of all of Alice's and Bob's sealed labs is encoded (assuming an argument à la Hayden--Preskill~\cite{hayden2007black}, where for an outsider the black hole dynamics is assumed to be unitary and rapidly mixing). 
Next, on the system containing Alice's decoded lab, Ursula ``undoes'' Alice's measurement by applying the inverse unitary $U_A^\dagger$ and measures $S_A$ in the $\ket{\pm}=\sfrac{1}{\sqrt{2}} (\ket{0}\pm\ket{1})$ basis with outcome $u \in \{\pm\}$.\footnote{One could also allow Alice (and Bob) to perform their measurement inside the black hole, and in that case question whether Ursula would still have to apply $U_A^\dagger$. I thank Eric Cavalcanti for this comment. This protocol may be more subtle, as it also involves the question of how Alice's (and Bob's) measurement inside is modeled from the outside.} Wigner acts analogously on Bob's decoded lab system, obtaining $w \in \{\pm\}$. 
Alice and Bob, once inside the black hole, try to send their outcomes $a,b$ towards Ursula and Wigner who fall into the black hole after having performed their measurements.\footnote{Note that Ursula and Wigner must decode the Hawking radiation at locations where they can still fall into the (evaporating) black hole, and thus cannot do this at late times near null or timelike future infinity. We comment more upon this in \Cref{sec:implications}.}
The protocol can be repeated many times.

\vspace{0.5cm}

Upon obtaining $u=-,w=-$, which has a nonzero probability of occurring, Ursula and Wigner reason as follows:
\begin{itemize}
    \item $u=-$ implies $b = 1$ as $\langle - |_{S_A} \langle 0|_{S_B}  \ket{\psi_0}_{S_A S_B} = 0$,
    \item $w=-$ implies $a = 0$ as $\langle 1|_{S_A} \bra{-}_{S_B} \ket{\psi_0}_{S_A S_B} = 0$,
    \item but $a=0,b=1$ cannot occur as $\langle 01|_{S_A S_B} \ket{\psi_0}_{S_A S_B}=0$,
\end{itemize} and thus a contradiction is obtained. This contradiction becomes operational if Alice and Bob, once inside the black hole, succeed in sending their outcomes and predictions to Ursula and Wigner. We could also have formulated the protocol by having Alice and Bob reason about the outcomes of Ursula and Wigner.

\vspace{0.5cm}

To explain in some more detail how the probabilities above are obtained, let us focus on the first, Ursula's claim about Bob's outcome $b$. 
 We start with the initial Hardy state $\ket{\psi_0}$ of \cref{eq:hardy} for the systems $S_A \otimes S_B$, and Alice's and Bob's memory initialized in some states, the $\ket{0}$ states for example, \begin{equation} \label{eq:Psi_0} \ket{\Psi_0} := \ket{\psi_0}_{S_A S_B} \ket{0}_A \ket{0}_B. \end{equation} 
 Alice and Bob perform their measurements, which are modeled unitarily as $U_A$ and $U_B$ acting on \cref{eq:Psi_0}, in this case CNOT gates with controls on $S_A$ and $S_B$, as explained in \Cref{sec:WF}, obtaining $U_A U_B \ket{\Psi_0}$. 
 
We next make explicit the idealized recovery assumption used in the protocol. Let $L=S_AAS_BB$ denote the relevant joint laboratory degrees of freedom of Alice and Bob, let $B_0$ denote the black hole before the relevant scrambling and evaporation, let $E$ denote previously emitted radiation, let $R$ denote the subsequently emitted radiation accessible to Ursula and Wigner, and let $B'$ denote the remaining black hole. 
 From the outsider's point of view (Ursula's and Wigner's point of view), Alice's and Bob's labs get scrambled by the black hole into Hawking radiation by evaporation, implemented by a unitary as 
\begin{equation}
V_{\mathrm{BH}}:
\mathcal H_L\otimes\mathcal H_{B_0}
\longrightarrow
\mathcal H_R\otimes\mathcal H_{B'} .
\end{equation}
We assume that, idealizing the recovery as exact on the relevant laboratory code subspace $L$, there exists a coherent decoding isometry
\begin{equation}
W_{\mathrm{dec}}:
\mathcal H_R\otimes\mathcal H_E
\longrightarrow
\mathcal H_{\widetilde L}\otimes\mathcal H_J
\end{equation}
that decodes the relevant information from the radiation $RE$, i.e. 
\begin{equation}
\label{eq:coherent_recovery}
\begin{split}
(W_{\mathrm{dec}}\otimes\mathbb 1_{B'})
(V_{\mathrm{BH}}\otimes\mathbb 1_E)
\ket{\phi}_L\ket{\Omega}_{B_0 E}
 =
\ket{\phi}_{\widetilde L}\ket{\chi}_{JB'}
\end{split}
\end{equation}
for every state $\ket{\phi}_L$ in that code subspace $L$. Here, $\widetilde L$ denotes the reconstructed laboratory, $J$ denotes unused output degrees of freedom of the decoder, and the state $\ket{\chi}_{JB'}$ is independent of $\ket{\phi}_L$. We henceforth identify the reconstructed laboratory degrees of freedom $\widetilde L$ with the corresponding original labels $L$.
 
 We consider now the case where Bob obtains $b=0$, giving the state
 \begin{equation} \label{eq:Psi_0_step2}
 \begin{split}
    & \ket{0}_{B} \bra{0}_{B} U_A U_B \ket{\Psi_0} \\ &= \ket{00}_{S_B B} \bra{0}_{S_B} U_A \ket{\psi_0}_{S_A S_B} \ket{0}_A 
 \end{split}
 \end{equation} 
 up to normalization, where we used that \begin{equation} \ket{0}_B \bra{0}_B U_B \ket{0}_B = \ket{00}_{S_B B} \bra{0}_{S_B}.\end{equation} 
 Next, Alice's and Bob's labs get scrambled by $V_{BH}$, and Ursula and Wigner apply the decoding map $W_\mathrm{dec}$ to the radiation (which includes both the early and late radiation $E$ and $R$), and thus obtain \cref{eq:Psi_0_step2} after decoding by \cref{eq:coherent_recovery}.
 Ursula next undoes Alice's measurement by applying $U_A^\dagger$ and performs her measurement, where we consider the case where her outcome is $u=-$, giving \begin{equation} \begin{split}
 \label{eq:Psi_final_prob0} & \ket{00}_{S_B B} \bra{-}_{S_A} U_A^\dagger \bra{0}_{S_B} U_A  \ket{\psi_0}_{S_A S_B} \ket{0}_A \\ &= \ket{00}_{S_B B} \bra{-}_{S_A} \bra{0}_{S_B} \ket{\psi_0}_{S_A S_B} \ket{0}_A  \\ &= 0,    
 \end{split} \end{equation} using that $U_A^\dagger U_A = \mathbb{1}$ and $\bra{-}_{S_A} \bra{0}_{S_B}  \ket{\psi_0}_{S_A S_B} =0$, resulting in a probability 0, $p(u=-,b=0)=0$. 
 The other probabilities used in Ursula's and Wigner's reasoning are obtained analogously.

\vspace{0.5cm}

We have thus proved the following no-go theorem.
\begin{theorem} \label{th:theorem_FR_BH}
    If a superobserver can decode the Hawking radiation and coherently reconstruct the quantum state of an infalling observer, and, when falling in, still receive a short message from that observer, sent inside the black hole and assumed to be unaltered by actions outside, an operational contradiction is found with quantum theory's predictions.
\end{theorem}

Compared to usual EWF arguments, here all of the outcomes $a,b,u,w$ seem available on a single spatial slice, as can be seen on \Cref{fig:radiation_mirror_timelike_path}(b), and \Cref{th:theorem_FR_BH} does not require the assumption of absolute outcomes of performed measurements: 
it involves only empirical correlations accessible to a single observer, made possible by assuming the interior records remain `unaltered by actions outside'.

\vspace{0.5cm}

If, as suggested in literature on the black hole cloning paradox (see \Cref{sec:BHcloning}), an observer catching the Hawking radiation from an infalling observer $A$ cannot obtain any message inside from $A$ due to time constraints, and so Ursula and Wigner are unable to obtain the outcomes of Alice and Bob inside, then we can still use our protocol to derive another no-go theorem. 
We keep the assumption, that physical records inside the black hole remain unaltered by operations outside, referred to below as $(Ci)$, and formalize the requirement that quantum theory's predictions remain valid for its use by observers through the assumptions $(Cii)$ and $(Ciii)$.

\begin{theorem} \label{th:nogo_EWFcloning_notmeet}
    If a superobserver can decode the Hawking radiation and coherently reconstruct the quantum state of an infalling observer, a contradiction is found if (Ci) the outcome records of an observer inside the black hole remain untouched by operations outside, (Cii) observers can make valid Born rule predictions about measurement outcomes (without actually obtaining those outcomes, but where those outcome records remain untouched) and (Ciii) observers who can communicate and meet have to agree on their predictions.
\end{theorem}

To clarify, the content of these theorems is not that a clone exists (which is the cloning paradox~\cite{susskind1994gedanken,hayden2007black}) but that the clone now produces a contradiction at the level of observers' operational probability predictions, rather than merely an incompatibility between two clones within quantum theory.

\subsection{Do the outcome records of observers inside get altered by operations outside?}

In the protocol above, the correlation $p(u,w)$ can be obtained experimentally by Ursula and Wigner, who remain outside the black hole, and the correlation $p(a,b)$ by Alice and Bob who fall into the black hole together.
However, if Alice and Bob fail to send a message with their outcomes inside the black hole to Ursula and Wigner, the correlations $p(u,b),p(a,w)$ are not verified experimentally as no single agent obtains both $u,b$, for example, which led to \Cref{th:nogo_EWFcloning_notmeet}.
So, in this case, why does Ursula argue that the joint correlations $p(u,b)$ of her prediction are those she would obtain if Bob had remained outside the black hole to compare her outcome to $b$? (Similarly for Wigner.) 

Similarly to the Local Friendliness no-go theorem~\cite{bong2020strong}, one can try to verify this prediction in certain rounds of the protocol, by allowing Alice to make a choice $x=0,1$ whether to leave her outcome outside the black hole or not (similarly for Bob with a choice $y$). 
In rounds where $x=0$ she leaves a copy of her outcome $a_O$ outside the black hole and the empirical correlations $p(a_O,w)$ are obtained outside the black hole. 
If Wigner can also access Alice's outcome $a$ inside, he finds that $p(a_O,w,a|x=0,y=1) = \delta_{a_O,a} p(a_O,w|x=0,y=1)$,  confirming consistency between inside and outside outcomes. 
Other choices of $x,y$ test other statements.
Implementing this idea in our first protocol we arrive at a Local Friendliness-style no-go theorem where an assumption of Local Agency is used to argue that all these statements remain true regardless of the choices $x,y$. 

An alternative way of motivating the validity of the quantum correlations $p(a,w),p(u,b)$ is to note that if Bob would fall in the black hole much earlier than Alice, then Wigner could decode Bob's Hawking radiation and perform his measurement with outcome $w$, while Alice remains outside still, so that they can verify the correlation $p(a,w)$. An analogous argument holds for Alice falling in earlier and the verification of $p(u,b)$.

\vspace{0.5cm}

The arguments above suggest that the correlations $p(a,w),p(u,b),p(a,b),p(u,w)$ are indeed those predicted by quantum theory as explained in \cref{eq:Psi_final_prob0}, if one can assign outcome values to the measurements of Alice and Bob who fell into the black hole. 
A possibility is that Ursula's and Wigner's actions erase Alice's and Bob's physical outcome records as in Wigner's friend scenarios.
However, semiclassical causality arguments would argue that Ursula's and Wigner's actions outside the black hole cannot alter the experiences of Alice and Bob inside.
Nevertheless, refuting this assumption, referred to as $(Ci)$ in \Cref{th:nogo_EWFcloning_notmeet} and also present in the cloning paradox, 
still seems one of the most reasonable ways out if one believes in unitarity, compared to refuting assumptions $(Cii)$ or $(Ciii)$ of \Cref{th:nogo_EWFcloning_notmeet}. 
As we will discuss in \Cref{sec:subtleties}, subtleties regarding black hole puzzles for infalling observers can be argued to make this option more natural.

\subsection{Comparison with the cloning paradox} \label{sec:BHcloning}

The cloning paradox analysed by Susskind and Thorlacius~\cite{susskind1994gedanken} and Hayden and Preskill~\cite{hayden2007black} is similar in spirit to our protocol, but does not directly provide a logical contradiction on the level of operational probability predictions.
Our result rules out any logically consistent theory that can describe the viewpoints of the different observers if they can use quantum theory for their own predictions, assuming memories inside the black hole remain unaltered by actions on the outside radiation.

In the cloning paradox, Alice falls into the black hole while carrying a quantum system $A$. 
Bob catches the Hawking radiation required to reconstruct $A$'s state, and dives into the black hole.
Alice, once inside the black hole, sends her system $A$ (or quantum information about it) to Bob, who would observe a violation of quantum theory's no-cloning.
Hayden and Preskill~\cite{hayden2007black} suggested that Bob, who collects the Hawking radiation outside to reconstruct a system $A$'s state, can just never retrieve the system $A$ inside. 
Therefore, an operational violation of no-cloning seems protected.
Namely, for Alice to send her qubit $A$ to Bob, according to Hayden and Preskill, she must operate within a Planck time scale on her proper time to send information about her qubit to Bob for it to reach him before he hits the singularity\footnote{This is assuming the causal structure of the Schwarzschild black hole, and need not be the case for the classical causal structure of the rotating and charged black holes. However, the actual (quantum) interior of the latter is likely quite different from the classical one~\cite{zilberman2022quantum,juarez2022quantumkicks,Ju_rez_Aubry_2015,juarez-aubry_quantum_2023,Lanir_2019,hollands2020quantum,Hollands_2020,Papadodimas_2020,cardoso2018quasi,Dias_2019,Dias_2019strong,Emparan_2020,mcmaken2023,zilberman2020quantumfluxes}.}. But this means that Alice must send her information in super-Planckian frequencies, in a strong gravity regime, likely much different from low-energy gravity. 
The same argument against an actual operational contradiction holds for the protocol of \Cref{sec:FR_BH_protocol}.
Nevertheless, the existence of these two copies in the cloning paradox remains problematic if one believes both views should be described by one single quantum theory.
Similarly, in the protocol above, even if Ursula and Wigner can never get any messages from Alice and Bob inside the black hole, a puzzle remains, as stated in \cref{th:nogo_EWFcloning_notmeet}.
In analogy with \Cref{th:nogo_EWFcloning_notmeet}, we could phrase the no-cloning theorem as: 
\begin{quote}
    If an observer can reconstruct a system $S$'s quantum state from the Hawking radiation of an infalling quantum system, a violation of no-cloning is found if $(Ci)$ the state of the quantum system $S$ that fell into the black hole remains coherent and unaltered by operations outside, and $(iv)$ quantum theory can describe the physics of the whole spacetime, including both the inside and outside of the black hole.
\end{quote}
One could propose to refute $(iv)$ and postulate a more general theory describing the whole spacetime, while requiring that it reproduce all operational predictions of quantum theory.
In the case where an operational violation of cloning would be found, our protocol and \Cref{th:theorem_FR_BH} above show that this is impossible if no single observer can falsify quantum theory's operational predictions. 
If no such operational violation is found, we still have the no-go result of \Cref{th:nogo_EWFcloning_notmeet}, under the additional assumption that the generalized theory should reproduce quantum theory's predictions for observers, formalized as assumptions $(Cii)$ and $(Ciii)$ in \Cref{th:nogo_EWFcloning_notmeet}.
This shows that the assumptions going into the black hole cloning paradox remain problematic even for such generalized theory if we believe in quantum theory's predictions.
We discuss these paradoxes further in \Cref{sec:implications}.

\section{Relating to literature} \label{sec:literature}

As an important part of this work concerns critically assessing standard black hole assumptions, we first recap the firewall paradox in \Cref{sec:firewall_recap}, to critically assess its assumptions in \Cref{sec:implications}. 
Next, we relate to the work of Hausmann and Renner on analogies between black hole puzzles and Wigner's friend~\cite{hausmann2025firewallparadoxwignersfriend} in \Cref{sec:HR}. 
We end this section with relating to literature on black hole puzzles and their resolutions in \Cref{sec:literature_BHpuzzles}.

\subsection{A recap of the firewall argument} \label{sec:firewall_recap}

\textbf{The firewall argument.} We follow the presentation of Ref.~\cite{hausmann2025firewallparadoxwignersfriend}. An infalling referee Bob is given a qubit $Q_B$ right outside an old black hole (i.e. past the Page time), being part of the late-time Hawking radiation. Two observers, Alice and Charlie, starting outside and inside the black hole, respectively, collaborate to provide two predictions $(P,\overline{P})$ to the referee Bob, for when he measures his qubit $Q_B$ in the $Z$- and $X$-bases. Alice and Charlie win if the referee's (Bob's) test succeeds; this is called the \textit{complementarity game}. Charlie, inside the black hole, can distill a qubit $Q_C$ (approximately) maximally entangled with $Q_B$ by the assumption of high entanglement across the horizon~\cite{braunstein2013better,reeh1961bemerkungen,eisert2010colloquium,Witten_2018,Hollands_2018,harlow2016jerusalem,mathur2009information} (assuming semiclassical gravity), and Alice, holding the early-time Hawking radiation outside the black hole, can distill a qubit $Q_A$ (approximately) maximally entangled with $Q_B$ by Page's argument~\cite{page1993information,page2013time}. Alice and Charlie measure their qubits $Q_A,Q_C$ in the $X$- and $Z$-bases, respectively, obtaining outcomes $a,c$. 
As Charlie's qubit is maximally entangled with $Q_B$, he predicts the referee's outcome $b$ to be the same as his outcome $c$ when he measures his qubit in the $Z$-basis, and similarly for Alice predicting outcome $a$ for the referee when he measures in the $X$-basis. They send their predictions $(a,c)$ to the referee who falls into the black hole (to retrieve Charlie's message and outcome). The predictions $P=a,\overline{P}=c$ yield a winning probability of 1, violating the quantum bound for winning the complementarity game based on a violation of monogamy of entanglement.

If some observer can obtain all outcomes of Alice, Bob and Charlie, an operational violation of quantum theory's formalism is found in the firewall argument. Such operational violation is more plausible than in the black hole cloning paradox, as no observer needs to wait for the Hawking radiation of an infalling observer to have come out. 
In fact, in typical treatments of the firewall argument, Alice, Charlie and the referee Bob are played by a single observer; this is possible if Alice's distilling operation can be performed sufficiently rapidly so she can still obtain Charlie's mode inside the black hole.
A potential problem in both of these paradoxes lies in the complexity of Alice's operation of distilling the qubit $Q_A$ from the early radiation, which could require a time exponentially long in the black hole entropy~\cite{harlow_quantum_2013,harlow2016jerusalem,aaronson_complexity_2016} compared to its polynomial evaporation time. 
However, with exponentially long preprocessing time such long decoding time could be avoided, as suggested in a toy black hole representation in Ref.~\cite{Oppenheim_2014}.

\vspace{0.5cm}

\textbf{The firewall paradox as a no-go theorem.} The firewall argument can be stated as a no-go theorem in the following way.

\begin{theorem}[Firewall paradox] \label{th:firewall} The firewall argument shows that the following four assumptions are incompatible:

    (Fi) the early radiation is maximally entangled with the late radiation---motivated by unitarity and the Page curve---and (Fii) an observer can distill a qubit $Q_A$ of the early radiation maximally entangled with a qubit mode $Q_B$ of the late radiation that is just outside the horizon; 
    
    (Fiii) the late radiation modes just outside the horizon are highly entangled with modes inside the black hole such that inside one can distill a qubit $Q_C$ maximally entangled with $Q_B$---motivated by the equivalence principle and high entanglement across the horizon as in QFT on curved spacetime; 

    (iv) quantum theory can describe the whole evolution comprising both the inside and outside of the black hole (including all radiation). 
\end{theorem}

One way to evade this no-go theorem would be to alter assumption $(iv)$\footnote{In typical treatments of the firewall and cloning paradoxes, assumption $(iv)$ is not explicitly stated, but implicitly assumed.}, by considering a more general theory than quantum theory that describes the evolution of both the inside and outside of the black hole. 
This route will not work if we want such theory to obey no-signalling. 
Namely, one can introduce measurement choices $x,y$ for Charlie and the referee Bob in the firewall paradox, and use the monogamy of entanglement violation to derive a contradiction of $(Fi)-(Fiii)$ with no-signalling (as defined in the semiclassical causal structure), as we prove in Appendix B. 
One can also combine the firewall and extended Wigner's friend arguments into a single unified scenario to derive a contradiction as in \Cref{th:nogo_EWFcloning_notmeet,th:theorem_FR_BH}; we refer the interested reader to Ref.~\cite[Chapter 8.3]{walleghem2026thesis}.

\subsection{Relating to the work by Hausmann and Renner} \label{sec:HR}

Hausmann and Renner (HR)~\cite{hausmann2025firewallparadoxwignersfriend} describe two Wigner's friend protocols and two protocols involving a black hole, and draw analogies between Wigner's friend and black hole paradoxes.
We have extended their work by combining an extended Wigner's friend protocol~\cite{frauchiger2018quantum,bong2020strong,cavalcanti2021implications,haddara2022possibilistic,walleghem_refined_2024} with black hole physics into a unified argument.
HR's first black hole argument is similar to the cloning paradox, and can be seen as combining a single Wigner's friend scenario with the cloning paradox. 
Their second no-go theorem, which we have described in \Cref{sec:firewall_recap}, is a rephrasing of the firewall argument where the non-monogamy of entanglement is used to violate the quantum bound of the complementarity game.

\vspace{0.5cm}

\textbf{How Hausmann and Renner relate to Wigner's friend.} \Cref{th:theorem_FR_BH} assumes that infalling observers can still send a message to the observers catching their Hawking radiation who fall into the black hole later, an assumption that may be problematic as argued in the context of the black hole cloning paradox (see \Cref{sec:BHcloning}), while \Cref{th:nogo_EWFcloning_notmeet} derives a no-go theorem without such assumption, based on the same protocol.
In the firewall paradox, the possibility of one single observer bringing the problematic aspects inside and outside the black hole together is more plausible as no observer needs to wait outside the black hole to catch the Hawking radiation of an infalling observer before falling into the black hole.
Hausmann and Renner~\cite{hausmann2025firewallparadoxwignersfriend} present a no-go theorem based on the firewall paradox as paraphrased in \Cref{sec:firewall_recap}, but with assumptions phrased differently than the ones of \Cref{th:firewall}.
A key assumption in their no-go theorem is a consistency assumption ($\mathcal{C}$), which allows Alice, inside the black hole, to reason about Bob outside the black hole. 
We can paraphrase the consistency assumption ($\mathcal{C}$), also used in EWF arguments such as the FR paradox~\cite{frauchiger2018quantum}, as follows: \begin{quote}
    ($\mathcal{C}$): If an observer $B$, modeled classically by observer $A$ (i.e., above $A$'s Heisenberg cut), knows that a statement $s$ about some measurement outcome(s) is true from reasoning with a theory that $A$ accepts, and $A$ knows about this reasoning of $B$, then $s$ is true for $A$ as well.
\end{quote}
Notably, assumption ($\mathcal{C}$) is already violated in certain classical paradoxes~\cite{jones_thinking_2024}, and is thus perhaps too strong. 
Therefore, one could apply the same criticism to Hausmann and Renner's theorems~\cite[Theorem 3 and 5]{hausmann2025firewallparadoxwignersfriend}.
However, the FR paradox can be slightly rephrased such that only a weaker consistency assumption is required~\cite{walleghem_refined_2024}, which we refer to as ($\mathcal{C}'$), where observers $A$ and $B$ only need to take each other's statements to be true if they can communicate and physically meet, notably not violated in the aforementioned classical paradoxes~\cite{walleghem2024stunnedsleepingbeautyprince}. 
We use a similar condition to ($\mathcal{C}'$) in \Cref{th:nogo_EWFcloning_notmeet}, the combination of assumptions $(Cii)$ and $(Ciii)$ there.

In fact, in the firewall paradox formulation of \Cref{sec:firewall_recap} (per Hausmann and Renner~\cite{hausmann2025firewallparadoxwignersfriend}), the assumption ($\mathcal{C}'$) could be used instead of ($\mathcal{C}$) if Bob were to fall into the black hole to meet Alice.
However, Hausmann and Renner refrain from letting Bob fall into the black hole, as they model his reasoning while he remains outside the black hole, used in the assumptions that as an outsider he can model the Hawking radiation process as a typical unitary for the early radiation and the modes at the stretched horizon (which will constitute the late radiation). 
Assuming no-retrocausality, one could argue that Bob only needs to model this unitary up to the point where he could verify his claims by asking the referee holding the modes at the stretched horizon, to use ($\mathcal{C}'$) instead of ($\mathcal{C}$).
We will not discuss this matter in more detail here.

\vspace{0.5cm}

The refined version of the FR paradox that uses ($\mathcal{C}'$) instead of ($\mathcal{C}$)~\cite{walleghem_refined_2024}, as well as other EWF arguments~\cite{bong2020strong,cavalcanti2021implications,frauchiger2018quantum,brukner2018no}, can be used to argue against absolute, single-valued outcomes of performed measurements, in favor of relationalism~\cite{rovelli1996relational,rovelli2018space,cavalcanti2023consistency,walleghem_refined_2024,debrota2020respecting,ormrod2024quantum}, assuming the existence of superobservers.
However, the no-go theorems in this work serve another purpose. 
Naive treatments of the interior and exterior of the black hole suggest that all relevant outcomes coexist on a single spatial slice, and thus arguing that these outcomes would not coexist as in EWF paradoxes such as the (refined) FR paradox seems somewhat odd here. 
Therefore, these no-go theorems rather suggest that these naive treatments of black hole physics must be false.

\subsection{Literature on black hole puzzles} \label{sec:literature_BHpuzzles}

The resolution of the black hole information puzzle remains heavily debated and inconclusive, but much work has been done since, hinting that information is preserved. 
Expectations that black hole evaporation is a unitary process come from AdS/CFT~\cite{maldacena1999large,harlow2016jerusalem}, and more recently from the proposed existence of `entanglement islands' inside the black hole giving a Page curve for the radiation~\cite{penington2020entanglement,almheiri2020page,almheiri2019entropy,wang2024entanglement,Hartman_2020,wang2021page,Hashimoto_2020,almheiri2020entanglementislandsinhigher,kibe2022holographic,chen2020quantumextremalislandseasyI,chen2020quantumextremalislandseasyII,bousso_islands_2023,almheiri2023islandsoutsidehorizon,Krishnan_2021,geng2021information,geng2026seeingpagecurvesislands}.\footnote{Concerns have been expressed that such Page curve from islands only arises in massive gravity~\cite{geng2022inconsistency,Geng_2020,raju2022failure,geng2025revisitingrecentprogresskarchrandall,geng2021information,geng2025makingcasemassiveislands}; in turn, a reply can be found in Ref.~\cite{antonini2025apologiaislands}, and a reply to this reply in Ref.~\cite{geng2026seeingpagecurvesislands}. See also footnote \ref{fn:replica_subtility}.} 
Replica trick calculations of entanglement entropy in 1+1d Euclidean gravity~\cite{mertens2023solvable} suggest that additional saddle points —-- so-called replica wormholes~\cite{almheiri2020replica,penington2022replica,Goto_2021}---appear when entanglement islands form.
Such replica wormholes lack a clear interpretation, but have been proposed to relate to black hole hair~\cite{calmet2024blackholeinformationreplica,an2023replica}.
Unitarity has also been argued for from two-dimensional investigations~\cite{ashtekar_information_2008,ashtekar_surprises_2011,Ashtekar_2011,barenboim2024no}.

Another proposal for information recovery consists of stable or long-lived remnants, objects that look small from the outside but store an enormous amount of information in their interior~\cite{Chen_2015,ong2025caseblackholeremnants}. An example of such proposal is black-to-white hole tunneling~\cite{DAmbrosio:2020mut,Christodoulou:2016vny,rovelli2014planck,Bianchi:2018mml,BenAchour:2020gon,Rovelli:2024sjl,Haggard:2014rza,Han:2023wxg}. Arguments in favor of remnants argue that there is a notion of volume that is very large for old black holes~\cite{Christodoulou:2014yia,Christodoulou:2016tuu}. 
Remnants are, however, often argued to encounter problems such as infinite pair production~\cite{Giddings_1995,susskind1995troubleremnants,Giddings:1993km}\footnote{For counterarguments see~\cite{ong2025caseblackholeremnants,Chen_2015,Bianchi:2018mml}.}. A related idea consists of baby universes that form inside black holes~\cite{Parentani_1994,Easson_2001}. If a (Planck-scale) remnant would hold the information, there is no need for the Hawking radiation to purify (until the remnants potentially decays and spits out the information it holds in low-energetic radiation), and thus also no Page curve for the (pre-remnant) Hawking radiation, no cloning and firewall paradoxes (and thus also no black hole Wigner's friend arguments of this work). We will not discuss this option further here.

Proposals that change the structure or existence of the horizon include, next to firewalls, fuzzballs~\cite{Mathur_2005}, nonviolent nonlocality~\cite{giddings2013nonviolent}, gravastars as an alternative to black holes~\cite{mazur2004gravitational} and that evaporation may be halted earlier than expected leading to `dark stars', which, however, by itself may not yield a full resolution to the information puzzle~\cite{KAWAI_2013,Baccetti_2017,Baccetti_2018,Carballo_Rubio_2018,cardoso2019testing,Chen_2018}.

One recent idea regarding a physical mechanism for unitary black hole evaporation includes soft hair, low-energy gravitons dressing black hole spacetimes that could purify the Hawking radiation~\cite{hawking2016soft,strominger2018lecturesinfraredstructuregravity,carney_infrared_2017,carney_dressed_2018,strominger2017blackholeinformationrevisited,flanagan_order-unity_2021,calmet2022quantuma}. Soft hair and soft theorems~\cite{weinberg_infrared_1965} are deeply tied to (gravitational) memory~\cite{Zeldovich:1974gvh,christodoulou1991nonlinear,thorne1992gravitational,Bieri_2024,maibach2026balancefluxlawsgeneral} and asymptotic symmetries~\cite{bondi1962gravitational,sachs1962gravitational,compère2019advancedlecturesgeneralrelativity,flanagan_conserved_2017,compere_poincare_2020,compere_asymptotic_2023,satishchandran2019asymptotic,prabhu_infrared_2024} through the infrared triangle~\cite{strominger2018lecturesinfraredstructuregravity}, and indeed argued to be of crucial importance for infrared problems in massless scattering~\cite{prabhu_infrared_2022,satishchandran2019asymptotic,prabhu_infrared_2024}.
Related proposals are quantum hair from higher-order curvature corrections~\cite{calmet2022quantumb,calmet2023quantum,calmet2024blackholeinformationreplica,calmet2022quantuma} and black-hole gravitational condensates~\cite{dvali2011blackholesquantumnportrait},
and a black hole wavefunction, which can be seen as a form of hair, that spreads or branches out~\cite{Eyheralde_2017,Eyheralde_2020,brustein2013restoring,alberte2015density,flanagan_order-unity_2021,calmet2024blackholeinformationreplica,calmet2022quantuma}.

Despite these arguments for unitarity, a clear, robust physical understanding of black hole paradoxes for infalling observers such as the firewall and cloning paradoxes remains elusive. 
One proposal is black hole complementarity~\cite{Stephens_1994,susskind1993stretched,susskind1994gedanken,Lowe_1995,kiem1995black} where the inside of a black hole and its radiation outside are not independent, and can be motivated from nontrivial relations between infalling matter and outgoing modes by backreaction~\cite{kiem1995black}.
Complementarity is only a principle, for which the firewall paradox was originally presented as a challenge~\cite{almheiri2013black,susskind2013blackholecomplementarityharlowhayden}, and can be formalized in different ways for an actual theory of the interior~\cite{harlow2016jerusalem}. 
Modeling black hole evaporation through quantum circuits suggests the importance of computational complexity and quantum error correction~\cite{harlow_quantum_2013,aaronson_complexity_2016,susskind2016computational,kim_ghost_2020,yoshida_firewalls_2019,akers2022blackholeinteriornonisometric}, where it has been argued that computational complexity protects the semiclassical causal structure of spacetime~\cite{kim_ghost_2020}, i.e. computational complexity preventing someone from creating a `firewall' of horizon excitations by manipulating the Hawking radiation.

However, despite complexity~\cite{aaronson_complexity_2016,harlow_quantum_2013,harlow2016jerusalem,Oppenheim_2014,akers2022blackholeinteriornonisometric,kim_ghost_2020,susskind2016computational,akers2022blackholeinteriornonisometric} and waiting time~\cite{hayden2007black} arguments to avoid operational contradictions for the firewall and cloning paradoxes, quantum theory cannot give a single description of two quantum clones or a violation of monogamy of entanglement.
Specific proposals for the interior include 
state-dependence~\cite{papadodimas2014state,papadodimas2014black,Harlow_2014},
also argued for in a milder form in the paradigm of entanglement wedge reconstructions and islands~\cite{penington2020entanglement,Cotler_2019entanglement}, final state projection~\cite{Lloyd_2014,Bousso_2014}, strong complementarity~\cite{harlow2016jerusalem} and proposals that modify aspects around the horizon, either violently such as firewalls, or (potentially) more subtly through nonviolent nonlocality~\cite{Giddings_2013,giddings2013nonviolent} and string-theoretic fuzzballs~\cite{Mathur_2005}, see also Refs.~\cite{harlow2016jerusalem,cardoso2019testing} for more discussion on other proposed compact quantum objects.
Other proposals view the interior or spacetime as emergent concepts~\cite{concepcion2024complementaritydynamicalblackhole,dvali2011blackholesquantumnportrait}, with Ref.~\cite{concepcion2024complementaritydynamicalblackhole} shown to exhibit a form of complementarity and state-dependence.

As we will discuss in \Cref{sec:implications}, important subtleties for infalling observers may be superposed quantum geometries and issues of locality and gauge-invariance.

\vspace{0.5cm}

\section{Discussion} \label{sec:implications}

We have presented scenarios combining Wigner's friend and black hole physics into a unified argument, resulting in \Cref{th:theorem_FR_BH,th:nogo_EWFcloning_notmeet} that extend the black hole cloning paradox, and lifted the firewall paradox to a violation of no-signalling (cf. Appendix B). 
Specifically, the cloning paradox leaves room for a generalized theory that describes the interior and exterior of the black hole to evade the no-cloning argument while not giving up on the standard black hole assumptions without violating quantum theory's predictions. 
With the scenarios in this work, we close this loophole, and show that no such theory exists if observers cannot witness an experimental violation of quantum theory's predictions.
Consequently, the problem must lie in other assumptions of the firewall and cloning paradoxes to prevent such violations.

In \Cref{sec:literature_BHpuzzles} we have argued that there are three broad classes of answers to the black hole information puzzle, (i) information loss, considered unlikely nowadays, (ii) remnant-like objects holding the  information, 
and (iii) aspects that were overlooked in Hawking's calculations, such as soft hair, superposed geometries, replica wormholes and islands, for example. 
It is within (iii) that paradoxes for infalling observers such as the cloning and firewall paradoxes take place.
In the remainder of this section, we will highlight subtleties in these paradoxes, and point toward further directions.
As outlined earlier, one possibility is to invoke modifications at the level of the horizon, or its status, such as firewalls, fuzzballs and gravastars. 
We will instead focus on subtleties that are present without necessarily invoking such new structures, and briefly discuss their implications.

\subsection{Subtleties in black hole paradoxes} \label{sec:subtleties}

\textit{a. Conservation laws and superposed geometries.}\label{sec:conservation_laws_superp_geometries} First, by conservation laws~\cite{compere_poincare_2020,flanagan_conserved_2017}, one expects an old black hole to be highly superposed, for example with a highly superposed mass~\cite{flanagan_infrared_2021}, also advocated for in Refs.~\cite{page1980is,Bao_2018,calmet2022brief,akil2025quantumsuperpositionblackhole,pasterski_hps_2021,Hutchinson_2016}.
Moreover, a black hole's soft hair in 4d asymptotically flat spacetime implies that the classical vacuum is highly degenerate~\cite{strominger_gravitational_2016,strominger2018lecturesinfraredstructuregravity}.
Therefore, drawing Penrose diagrams\footnote{Additionally, it is unclear how to continue a Penrose diagram post-evaporation as the near-singularity region is one of strong quantum gravity, see for example \Cref{fig:Penrose_post_evap} for two different proposals. 
This is why in later Penrose diagrams of \Cref{fig:critique_cloning_firewall} we will add a question mark `$?$' for how exactly the post-evaporation spacetime should be pictured.} with single-spacetime slices into the bulk of an old black hole may be problematic, and observers falling into the black hole can get entangled with its properties through the gravitational interaction, or even measure them.\footnote{One possibility would be to construct an effective spacetime metric from the correlators in the superposed spacetimes~\cite{Kempf_2021,saravani2016spacetime,foo2022quantum}; this requires further exploration and may not necessarily yield the standard black hole metric. See also Ref.~\cite{akil2023semiclassical} for semiclassical effective metrics as averages over superposed metrics.}

\vspace{0.5cm}

\textit{b. Locality and gauge invariance.}\label{sec:locality_gauge_invariance} Secondly, the notions of locality and gauge invariance\footnote{I thank David Mattingly for many interesting discussions and exchanges on these issues.}  in (quantum) gravity are subtle, and may have substantive implications for black hole (infaller) paradoxes~\cite{JACOBSON_2013,Jacobson_2019,Papadodimas:2015xma,papadodimas2014state,JACOBSON_2013,geng2026mechanisminformationencodingislands,raju2019toy,Chakraborty_2022}.
To make a localized observable gauge-invariant, one can use relational observables or gravitational dressings~\cite{Rovelli:1990ph,Brown_1995,Giddings:2019hjc,Giddings:2022hba,Giddings:2025xym,Giddings:2025bkp,Donnelly:2016rvo,Donnelly:2018nbv,Giddings_2006,goeller2022diffeomorphisminvariantobservablesdynamicalframes,Jacobson_2019,Giddings_2019,Giddings:2024qcf,JACOBSON_2013,Papadodimas:2015xma,papadodimas2014state,goeller2022diffeomorphisminvariantobservablesdynamicalframes,geng2022inconsistency}, with nontrivial effects regarding spacetime locality. 
Such dressings have been argued to be of crucial importance for the mechanism behind islands~\cite{geng2026mechanisminformationencodingislands}.
A related problem is the impossibility to locally factorize the Hilbert spaces of the outside radiation and inside of the black hole in the bulk in quantum gravity~\cite{JACOBSON_2013,Jacobson_2019,raju2022failure}.

\vspace{0.5cm}

\textit{c. Where is a Page curve obtained? At future null infinity?}\label{sec:where_page_curve}
Related to the previous point, one could argue that unitarity from the radiation and the Page curve\footnote{\label{fn:replica_subtility}If gravity is holographic~\cite{hooft2009dimensionalreductionquantumgravity,Susskind_1995,witten1998antisitterspaceholography,polchinski1999smatricesadsspacetime,susskind1999holography,maldacena1999large,giddings2000flat,Aharony_2000,de_Boer_2003,Dappiaggi_2004,Arcioni_2004,Arcioni_2003,DAPPIAGGI_2006,papadodimas2014state,Marolf_2009,Marolf_2013,JACOBSON_2013,Giddings_2018gauge,Giddings_2020,Donnay_2022carrollian,donnay_celestial_2023,pasterski2023chaptercelestialholography,laddha_2021,raju2019toy,raju2022failure,Chowdhury_2022,Araujo_Regado_2023}, one might have to couple the boundary, such as the AdS one in recent replica-trick calculations~\cite{almheiri2019entropy,almheiri2020replica,almheiri2021entropy,penington2020entanglement,penington2022replica,geng2021information}, to a non-gravitational bath to store the radiation and obtain a Page curve, argued to lead to massive gravitons through a Higgs mechanism in Refs.~\cite{raju2022failure,geng2022inconsistency,geng2025revisitingrecentprogresskarchrandall,Geng_2020,geng2025makingcasemassiveislands,geng2026mechanisminformationencodingislands}. Another possibility is to couple the black hole to a gravitational bath, and define the black hole and bath system (the radiation) through a non-geometric split~\cite{Uhlemann2022information,Uhlemann_2021,geng2021information}.} are only obtained near or at future (null) infinity in asymptotically flat spacetimes, which is also where the radiation was collected in an asymptotically flat study of islands~\cite{Hartman_2020}, and that gauge-invariantly identifying the near horizon modes and their evolution to form the late radiation arriving at future (null) infinity has nontrivial implications.
Closer to an old black hole, one can still have the ability to measure or get entangled with the (superposed) gravitational field; isolating the radiation and decoding information from it would involve a nontrivial construction in the bulk, with potentially backreaction of that construction being non-negligible~\cite{Hui_2014}, and one has the failure of factorization per \hyperref[sec:locality_gauge_invariance]{$b$} there;  
see also \Cref{fig:critique_cloning_firewall}.

Therefore, regarding the cloning paradox, even if the Page curve suggests that one would only require a few qubits from an old black hole to reconstruct the quantum state of a system $S$ that just fell into the black hole~\cite{hayden2007black}, a possibility could be that one might only recover those radiation qubits at or near future (null) infinity, with no `nice' slices~\cite{preskill1992black,Lowe_1995,giddings2006black,Puhm_2017,Giddings_2006,polchinski1995stringtheoryblackhole,harlow2016jerusalem} hugging future null infinity until late times in an evaporating black hole spacetime while running into the black hole interior containing the system $S$ that fell in, as in \Cref{fig:critique_cloning_firewall}.
Additionally, potential issues of nonlocal and non-adiabatic effects on nice slices have been expressed before~\cite{Giddings_2006,Lowe_1995,Puhm_2017}.

If information recovery and a Page curve are only to be expected at or near future (null) infinity $I^+$, a possibility could be that operations on the radiation do change the memory of an infalling observer, but, with the same causality of Wigner's friend; only when that memory arrives at future (null) infinity encoded in the radiation.
Such causality picture, with experiences during infall taking place earlier than the collection of that information at infinity, was advocated for recently in Ref.~\cite{concepcion2024complementaritydynamicalblackhole}.
Naturally, questions of the full experience of an infalling observer are still limited by the black hole singularity, and dependent upon its potential resolution. 
Speculations aside, the issues highlighted in this section call for an operational view on causality in quantum gravity~\cite{Hardy_2007,oreshkov2012quantum,chiribella2013quantum,castroruiz2018dynamics,Parker_2022}.

\begin{figure}[h]
         \centering
\tikzset{every picture/.style={line width=0.75pt}} 

\begin{tikzpicture}[x=0.75pt,y=0.75pt,yscale=-1,xscale=1]

\draw [color={rgb, 255:red, 0; green, 0; blue, 0 }  ,draw opacity=1 ][line width=0.75]    (122.81,196.25) -- (198.52,135.6) ;
\draw    (121.92,134.48) .. controls (123.61,132.84) and (125.28,132.86) .. (126.92,134.55) .. controls (128.56,136.24) and (130.23,136.27) .. (131.92,134.63) .. controls (133.61,132.99) and (135.27,133.01) .. (136.92,134.7) .. controls (138.57,136.39) and (140.23,136.41) .. (141.92,134.77) .. controls (143.61,133.13) and (145.27,133.15) .. (146.92,134.84) .. controls (148.56,136.53) and (150.23,136.56) .. (151.92,134.92) .. controls (153.61,133.28) and (155.27,133.3) .. (156.92,134.99) .. controls (158.57,136.68) and (160.23,136.7) .. (161.92,135.06) .. controls (163.61,133.42) and (165.28,133.45) .. (166.92,135.14) .. controls (168.57,136.83) and (170.23,136.85) .. (171.92,135.21) .. controls (173.61,133.57) and (175.27,133.59) .. (176.92,135.28) .. controls (178.56,136.97) and (180.22,136.99) .. (181.91,135.35) .. controls (183.6,133.71) and (185.27,133.74) .. (186.91,135.43) .. controls (188.56,137.12) and (190.22,137.14) .. (191.91,135.5) .. controls (193.6,133.86) and (195.26,133.88) .. (196.91,135.57) -- (198.52,135.6) -- (198.52,135.6) ;
\draw [color={rgb, 255:red, 255; green, 255; blue, 255 }  ,draw opacity=1 ]   (186.97,133.81) .. controls (188.64,132.16) and (190.31,132.17) .. (191.97,133.84) .. controls (193.62,135.52) and (195.29,135.53) .. (196.97,133.88) .. controls (198.64,132.23) and (200.31,132.24) .. (201.97,133.91) .. controls (203.63,135.58) and (205.3,135.59) .. (206.97,133.94) .. controls (208.64,132.29) and (210.31,132.3) .. (211.97,133.97) .. controls (213.63,135.64) and (215.3,135.65) .. (216.97,134) .. controls (218.65,132.35) and (220.32,132.36) .. (221.97,134.04) .. controls (223.63,135.71) and (225.3,135.72) .. (226.97,134.07) .. controls (228.64,132.42) and (230.31,132.43) .. (231.97,134.1) .. controls (233.63,135.77) and (235.3,135.78) .. (236.97,134.13) .. controls (238.64,132.48) and (240.31,132.49) .. (241.96,134.16) .. controls (243.61,135.84) and (245.28,135.85) .. (246.96,134.2) -- (249.46,134.21) -- (249.46,134.21) ;
\draw [color={rgb, 255:red, 0; green, 0; blue, 0 }  ,draw opacity=1 ]   (122.51,242.49) -- (241.27,148.18) ;
\draw [color={rgb, 255:red, 208; green, 2; blue, 27 }  ,draw opacity=1 ][line width=1.5]    (122.23,134.48) -- (188.03,189.91) ;
\draw [color={rgb, 255:red, 0; green, 0; blue, 0 }  ,draw opacity=1 ]   (198.27,110.62) -- (241.27,148.18) ;
\draw [color={rgb, 255:red, 0; green, 0; blue, 0 }  ,draw opacity=1 ]   (122.51,242.49) -- (122.23,134.48) ;
\draw [color={rgb, 255:red, 0; green, 0; blue, 0 }  ,draw opacity=1 ]   (198.27,110.62) -- (198.52,135.6) ;
\draw [color={rgb, 255:red, 0; green, 0; blue, 0 }  ,draw opacity=1 ][line width=0.75]    (312.6,206.09) -- (367.44,160.39) ;
\draw [color={rgb, 255:red, 255; green, 255; blue, 255 }  ,draw opacity=1 ]   (375.01,144.15) .. controls (376.68,142.49) and (378.35,142.5) .. (380,144.18) .. controls (381.66,145.85) and (383.33,145.86) .. (385,144.21) .. controls (386.68,142.56) and (388.35,142.57) .. (390,144.25) .. controls (391.66,145.92) and (393.33,145.93) .. (395,144.28) .. controls (396.67,142.63) and (398.34,142.64) .. (400,144.31) .. controls (401.66,145.98) and (403.33,145.99) .. (405,144.34) .. controls (406.67,142.69) and (408.34,142.7) .. (410,144.37) .. controls (411.65,146.05) and (413.32,146.06) .. (415,144.41) .. controls (416.67,142.76) and (418.34,142.77) .. (420,144.44) .. controls (421.66,146.11) and (423.33,146.12) .. (425,144.47) .. controls (426.67,142.82) and (428.34,142.83) .. (430,144.5) .. controls (431.66,146.17) and (433.33,146.18) .. (435,144.53) -- (437.5,144.55) -- (437.5,144.55) ;
\draw [color={rgb, 255:red, 0; green, 0; blue, 0 }  ,draw opacity=1 ]   (312.9,253) -- (410.88,170.04) ;
\draw [color={rgb, 255:red, 208; green, 2; blue, 27 }  ,draw opacity=1 ][line width=1.5]    (311.19,160.13) -- (369.94,205.44) ;
\draw [color={rgb, 255:red, 0; green, 0; blue, 0 }  ,draw opacity=1 ]   (310.55,83.4) -- (410.88,170.04) ;
\draw [color={rgb, 255:red, 0; green, 0; blue, 0 }  ,draw opacity=1 ]   (312.9,253) -- (310.55,83.4) ;
\draw  [draw opacity=0][fill={rgb, 255:red, 208; green, 2; blue, 27 }  ,fill opacity=1 ] (309.44,159.71) .. controls (316.07,153.44) and (326.83,149.33) .. (339,149.27) .. controls (351.66,149.21) and (362.85,153.54) .. (369.49,160.18) -- cycle ;
\draw  [draw opacity=0][fill={rgb, 255:red, 208; green, 2; blue, 27 }  ,fill opacity=1 ] (367.98,159.78) .. controls (361.17,164) and (351.28,166.68) .. (340.26,166.72) .. controls (328.72,166.77) and (318.39,163.92) .. (311.51,159.41) -- cycle ;
\draw    (311.51,159.41) .. controls (313.19,157.75) and (314.86,157.76) .. (316.51,159.44) .. controls (318.17,161.11) and (319.84,161.12) .. (321.51,159.47) .. controls (323.19,157.82) and (324.86,157.83) .. (326.51,159.51) .. controls (328.17,161.18) and (329.84,161.19) .. (331.51,159.54) .. controls (333.18,157.89) and (334.85,157.9) .. (336.51,159.57) .. controls (338.17,161.24) and (339.84,161.25) .. (341.51,159.6) .. controls (343.19,157.95) and (344.86,157.96) .. (346.51,159.64) .. controls (348.17,161.31) and (349.84,161.32) .. (351.51,159.67) .. controls (353.18,158.02) and (354.85,158.03) .. (356.51,159.7) .. controls (358.16,161.38) and (359.83,161.39) .. (361.51,159.74) .. controls (363.18,158.09) and (364.85,158.1) .. (366.51,159.77) -- (367.98,159.78) -- (367.98,159.78) ;

\draw (99,102.83) node [anchor=north west][inner sep=0.75pt]   [align=left] {$\displaystyle ( a)$};
\draw (285.46,111.09) node [anchor=north west][inner sep=0.75pt]   [align=left] {$\displaystyle ( b)$};
\draw (324.64,151.63) node [anchor=north west][inner sep=0.75pt]  [font=\footnotesize,color={rgb, 255:red, 255; green, 255; blue, 255 }  ,opacity=1 ] [align=left] {$\displaystyle QG$};

\end{tikzpicture}
         \caption{To include discussion about evaporation of black holes and unitarity, the post-evaporation spacetime must be pictured as well. Two possibilities are (a): continuing past the singularity, and (b) the same continuation but for a nonsingular black hole; see also Refs.~\cite{Ashtekar_2005,hawking2016soft,Schindler_2020,Ashtekar_2020,Ashtekar_2025,hiscock1981models,hiscock1981parttwo,Hayward_2006,Ashtekar_2011,ashtekar_surprises_2011,Stephens_1994,barenboim2024no,Carballo-Rubio:2025fnc,almheiri2019entropy}. 
         }
         \label{fig:Penrose_post_evap}
\end{figure}

\vspace{0.5cm}

\textit{d. More subtleties and observer-dependencies in the firewall paradox.} \label{sec:more_subtleties_firewall} Fourth, regarding the firewall paradox, the appearance of superposed geometries for old black holes was already mentioned above in \hyperref[sec:conservation_laws_superp_geometries]{$a$}, conflicting with the assumption $(Fiii)$ in \Cref{th:firewall} of high entanglement across the horizon derived from QFT on a single spacetime, as well as the non-factorization regarding locality and gauge invariance per \hyperref[sec:locality_gauge_invariance]{$b$} for specifying the early radiation and near horizon (interior and exterior) modes.

Moreover, also in QFT, the experience of an infalling observer, and the entanglement they see, may be very different from an observer remaining outside (for longer)~\cite{Barbado_2011,Barbado_2016}.
Consequently, the descriptions of the black hole's evolution and entanglement profile for different observers could be quite distinct~\cite{Fuentes_Schuller_2005,Alsing_2012,devuyst2025gravitationalentropyobserverdependent}, with entanglement known to be a frame-dependent notion~\cite{giacomini2019quantum} and potential subtleties regarding the equivalence principle for infalling observers~\cite{Barbado_2016,Barbado_2011,smerlak2013new,ng2022little,preciado2024more}.
As mentioned in the previous point \hyperref[sec:where_page_curve]{\textit{c}}, assumption $(Fii)$ of the firewall paradox as presented in \Cref{th:firewall} can also be questioned, with late-time radiation modes only required to purify the early radiation when arriving at infinity, and need not necessarily be trivially identified with near-horizon modes.

\vspace{0.5cm}

\textit{e. Infaller's backreaction and initial states.}\label{sec:infaller_backreaction} Fifth, the backreaction of an infalling object or observer is a nontrivial effect~\cite{Dray:1985yt,hooft1987strings,kiem1995black,Stephens_1994,yoshida_firewalls_2019,gaddam_soft_2024,penington2020entanglement,Ferrari:1988cc,Sfetsos_1995,shenker2014black,Betzios:2016yaq,Gaddam:2021zka,BenTov:2017kyf,polchinski2015chaosblackholesmatrix,pasterski_hps_2021}. 
Backreaction due to infalling objects and aspects of ultrarelativistic scattering have been investigated using shockwaves~\cite{HOOFT198761,hooft1987strings,Dray:1985yt,Ferrari:1988cc,Sfetsos_1995,BenTov:2017kyf,Verlinde_1992,polchinski2015chaosblackholesmatrix,shenker2014black}, but can also be probed using (black hole) eikonal approximations to scattering~\cite{gaddam_soft_2024,Gaddam:2020rxb,Gaddam:2021zka,divecchia2024gravitationaleikonalparticlestring,Amati_2008,Veneziano_2008,Veneziano_2008II,Marchesini_2008high,Kabat_1992,Di_Vecchia_2020,Ciafaloni_2008,Dvali_2015,Addazi_2017,ciafaloni2019infrared,banks1999modelhighenergyscattering}. 

Furthermore, 
maintaining the coherence of the quantum information encoded in an infalling system as observed by a co-infalling observer, a crucial assumption in the cloning paradox for example, requires careful shielding\footnote{I thank David Mattingly for interesting discussions on this matter.} of that quantum information to prevent leakage and decoherence. 
This is complicated by the standard assumption that gravity does not allow for shielding~\cite{bertolami2006generaltheoryrelativitysurvive,savrov2012gravitational,Dukehart:2024bsn}, implying that such quantum information should be encoded in other degrees of freedom.
Recent arguments of decoherence of superpositions due to (soft) radiation leaking into horizons or null infinity demonstrate such complications~\cite{Danielson:2022tdw,Danielson:2024yru,Belenchia_2018,danielson2025horizonssoftquantuminformation}.

Relatedly, a full treatment of the unitarity problem of black hole evolution requires specification of the initial \textit{quantum} state for black hole formation, where literature separates into collapsing shells~\cite{petr_kay_kuchar1992quantum,hajicek1992quantum,hajivcek2001unitary,louko_hamiltonian_1998,corichi2002quantum,vaz2007quantum,Vaz_2022,Eyheralde_2017,baccetti2019black,gaztanaga2025gravitational,berezin1997quantum,Malafarina_2017,ziprick2016polymer,rovelli2014planck}, high-energy collisions of two particles~\cite{Eardley_2002,giddings2004black,Hsu_2003,Veneziano_2008,divecchia2024gravitationaleikonalparticlestring,Choptuik_2010,sperhake_cardoso_2009cross,yoshiblackno2003}, holographic investigations of collapsing shells~\cite{Anous_2016} and colliding particles~\cite{Haehl_2023}, and Euclidean path integral preparations of black holes in two-dimensional gravity~\cite{mertens2023solvable,kourkoulou2017purestatessykmodel}.

\begin{figure}[]
         \centering
\includestandalone[width=0.5\textwidth]{figures/critique_cloning_firewall}
         \caption{A critique of the cloning paradox (a), with the state $\ket{\psi}$ cloned on a spatial surface, and the firewall paradox (c), with three systems violating monogamy of entanglement, the inside mode highly entangled with the mode just outside the horizon and the early radiation also highly entangled with the same late radiation mode just outside the horizon. Diagrams (b) and (d) give alternative viewpoints for the cloning and firewall paradox, respectively. Namely, in order to recover the information, it is only at or near $I^+$ that one must be able to reconstruct the state $\ket{\psi}$ of an infalling system from the radiation and potentially remaining gravitational degrees of freedom. Similarly, it is only at or near $I^+$ that the early radiation must be highly entangled with the late radiation. Furthermore, the experiences of infalling observers might be quite different as they get entangled with the superposed, backreacted geometry (in which a notion of single spatial slices may be questionable as well).}
         \label{fig:critique_cloning_firewall}
\end{figure}

\subsection{Implications and future directions} \label{sec:future_directions}
The considerations above raise several doubts upon typical assumptions in black hole paradoxes for infalling observers, which also underlie our no-go theorems. 
Specifically, the locality and gauge-invariance issues of \hyperref[sec:locality_gauge_invariance]{$b$}, further complicated by the superposed spacetime of an old black hole per \hyperref[sec:conservation_laws_superp_geometries]{$a$}, and the related problems of causality, and how and where information is recovered from the radiation per \hyperref[sec:where_page_curve]{$c$}, as well as the infaller's backreaction in \hyperref[sec:infaller_backreaction]{$e$}, question how independent the relevant interior and exterior operators in black hole paradoxes are. 
These elements thus question the validity of assumption $(Ci)$ in the cloning paradox and \Cref{th:theorem_FR_BH,th:nogo_EWFcloning_notmeet}, that quantum information `inside' remains unaltered by operations `outside' the black hole, with inside and outside referring to a semiclassical picture.

Moreover, superposed geometries and soft hair, non-factorization, the location of information retrieval and observer-dependencies as in \hyperref[sec:conservation_laws_superp_geometries]{$a$}, \hyperref[sec:locality_gauge_invariance]{$b$}, \hyperref[sec:where_page_curve]{$c$} and \hyperref[sec:more_subtleties_firewall]{$d$} present challenges for assumptions $(Fii)$ and $(Fiii)$ of the firewall paradox of \Cref{th:firewall}, namely the relevance of high entanglement across the horizon and the identification of late-time (purifying) radiation modes with near horizon modes.

Therefore, with the aim of progressing towards a better physical understanding, operational investigations of assumptions and explicit modeling of infalling quantum systems in the context of black hole puzzles are an interesting avenue, with potential for lab experiments in analogue gravity~\cite{barcelo2005analogue,unruh1995sonic,steinhauer2016observation,lahav2014realisation,drori2019observation,Steinhauer_2014,bose2025massive,Munoz2019observation,liberati2019information,chowdhury2022sachdev}. 
One such possibility would entail the use of quantum detectors, where theoretical investigations with detectors in (nonclassical) spacetimes have given us a rich plethora of results already~\cite{Barbado_2011,Barbado_2016,ng2022little,henderson2018harvesting,Ng_2014,bhattacharya_probing_2024,Smith_2014,Barbado_2011,Hodgkinson_2012,paczos2023hawking,henderson2018harvesting,Henderson_2020,Henderson_2022,Ju_rez_Aubry_2014,preciado2024more,wang2024singular,spadafora2024deepknottedblackhole,Foo_2020,PhysRevResearch.3.043056,foo2021entanglement,foo2023quantum,goel2024accelerateddetectorsuperposedspacetime,foo2022quantum,suryaatmadja2023signaturesrotatingblackholes,niermann2024particle,paczos2023hawking,Stritzelberger_2021,Hotta_2020,Stritzelberger_2020,Cong_2020,Ng_2018,henderson2018harvesting,Gallock_Yoshimura_2021,Mendez_Avalos_2022,Tjoa_2020,robbins2020entanglementamplificationrotatingblack,chakraborty2024entanglementharvestingquantumsuperposed,vsoda2022acceleration,pan2024gravity,chen2024quantum,chen2023quantum,foo2022quantum,paczos2023hawking}.
Furthermore, as black hole puzzles concern different observers' perspectives, questions of gauge invariance, dressings, locality and soft and edge modes, with moreover radiation leaking into or across (reference frames at) null infinity, the paradigm of quantum reference frames~\cite{hoehn_trinity_2021,giacomini2019quantum,giacomini_spacetime_2021,vanrietvelde_switching_2023,rovelli_quantum_1990,goeller2022diffeomorphisminvariantobservablesdynamicalframes,kabel2023quantum,kabel2024identification,Fewster_2024,ahmad2024quantum,Carrozza_2024,kabel2023quantum,Araujo_Regado_2025} may prove to be a valuable framework for further investigations of some of these subtleties.  

\vspace{0.5cm}

Finally, computational complexity and quantum error correction have been associated numerous times with black holes, holography and the decoding of Hawking radiation~\cite{harlow_quantum_2013,aaronson_complexity_2016,kim_ghost_2020,Oppenheim_2014,susskind2013blackholecomplementarityharlowhayden,susskind2016computational,brown2019pythonslunchgeometricobstructions,harlow2016jerusalem,akers2022blackholeinteriornonisometric,Brown_2017,BAIGUERA20261,engelhardt2023algebraicereprcomplexitytransfer,Belin_2022,almheiri2018holographicquantumerrorcorrection,Verlinde_2013,Hayden_2019,Almheiri_2015,Pastawski_2015}, but have not been probed in the context of Wigner's friend.  
Where are \textit{our} superobservers? Why can we trust our memories, for example for formulating physical laws? If Wigner can manipulate the brain of his friend, Wigner could implement memories and experiences that would lead the friend to different physical laws, or even nonsensical results.
We can imagine a similar situation when an observer, let's say a bee, falls into a fire. 
It burns to ashes, but we expect that when carefully keeping track of all relevant pieces of information, we can reconstruct the bee in the (classical) state right before it fell in. 
Now if we could perform such complex operation, then perhaps we could also manipulate its brain into thinking that right before it fell into the fire the world was filled with flying crocodiles. 
Perhaps emergence and computational complexity are undervalued concepts regarding consistency, Wigner's friend experiments and related (self-referential) puzzles such as Boltzmann brains~\cite{jones_thinking_2024,carroll2017boltzmannbrainsbad}.

\section*{Acknowledgements}
I thank David Maibach, Rui Soares Barbosa, Bernard Kay and Benito Juárez-Aubry for feedback on this manuscript.
I thank David Maibach, Benito Juárez-Aubry, Bernard Kay, Claudio Dappiaggi and Ricardo Faleiro for many interesting discussions on black holes and the information puzzle, and Časlav Brukner on causality and the firewall paradox.
I thank Lorenzo Catani for making me aware of the result of Hausmann and Renner, and Rui Soares Barbosa, Eric Cavalcanti, Matt Pusey, Stefan Weigert, Yìlè Yīng, David Schmid, Rafael Wagner and Lorenzo Catani for interesting discussions on Wigner's friend. 
Finally, I would like to thank David Mattingly for many interesting discussions and exchanges on gravitational shielding and gauge invariance.
I acknowledge support from the United Kingdom Engineering and Physical Sciences Research Council (EPSRC) through the DTP Studentship EP/W524657/1, and thank INL and its QLOC group in Braga, Portugal for the kind hospitality.

\vspace{0.5cm}

\section*{Appendix A: A short recap of the Frauchiger--Renner paradox and Local Friendliness no-go theorem} 

The protocols of the Frauchiger--Renner~\cite{frauchiger2018quantum} and Local Friendliness~\cite{bong2020strong,cavalcanti2021implications} both rely on quantum realisations of Bell nonlocality~\cite{bell1964einstein,Bell_1976,brunner_bell_2014}.
More specifically, the Frauchiger--Renner paradox transforms Hardy's example~\cite{hardy1993nonlocality} of a Bell scenario with two parties $A,B$ into a protocol with two superobserver-observer pairs. 
In Hardy's Bell scenario, each party can choose one out of two measurements to perform, while in the FR protocol the two measurements are performed in a single round by the corresponding superobserver-observer pair: one by the observer, and the other by their corresponding superobserver.

In the Frauchiger--Renner paradox, Alice and Bob share a Hardy state $\ket{\psi_0}_{S_A S_B} = \frac{1}{\sqrt{3}}(\ket{00}+\ket{10}+\ket{11})_{S_A S_B}$ and measure their system in the computational basis, modeled unitarily as $U_A,U_B$. Next Ursula and Wigner undo their measurements by applying $U_A^\dagger, U_B^\dagger$ and measure $S_A,S_B$ in the $X$-basis, obtaining outcomes $u,w$, respectively. 
Upon Ursula and Wigner obtaining $u=-,w=-$, Ursula reasons about which outcome Bob has, and reasons about what Bob predicts for Alice's outcome, and what Bob thinks Alice's prediction for Wigner is. All agents use quantum theory to formulate their prediction: from $u=-$ Ursula argues that $b=1$, and from $b=1$ she argues that Bob reasons that $a=0$, from which in turn Alice would argue that $w=+$.
From this chain of reasoning, Ursula finds a contradiction.

A stronger paradox can be found by having only classical observers (that is, observers who need not be modeled unitarily by anyone in the protocol) doing the reasoning~\cite{walleghem_refined_2024}.

In the Local Friendliness protocol, additional choices (also referred to as interventions or free choices in literature) $x,y$ are given to Ursula and Wigner whether to perform their measurement as just described or look at the outcome of Alice and Bob, respectively. 
In this way, by having Ursula look at Alice's outcome and Wigner perform his measurement, the correlation $a,w$ is empirically accessible after the protocol.
A contradiction is derived when assuming the Local Friendliness assumptions, i.e. the conjunction of absolute outcomes of performed measurements and a notion of locality named Local Agency, which allows only events in the future lightcone of an intervention to be correlated with that intervention. Local Agency can be seen as a stronger form of no-signalling~\cite{walleghem2025extendedwignersfriendnogo}, which is operational.

Both the FR and LF arguments can be constructed from other realizations of Bell nonlocality (and contextuality)~\cite{walleghem_refined_2024,haddara2022possibilistic,haddara2024local,walleghem2023extended,walleghem2024connecting}.

\section*{Appendix B: No-signalling firewall argument} \label{sec:strengthened_firewall}

Consider the following protocol, a slightly altered firewall paradox. 
Alice holds the early radiation, and distills a qubit (possibly precomputed using Oppenheim and Unruh's argument if its computational complexity would require so~\cite{Oppenheim_2014}), maximally entangled with the (late-radiation) mode $B$ that Bob holds near the horizon of an old black hole: $1/\sqrt{2}(\ket{00}+\ket{11})_{AB}$. By the firewall assumptions $(Fi)-(Fiii)$, the mode $B$ is also maximally entangled with Charlie's mode inside the black hole: $1/\sqrt{2}(\ket{01}-\ket{10})_{CB}$.
Bob chooses $y=0,1$ to measure his qubit $B$ in the $Z$ or $X$ basis. Alice always measures her qubit $A$ in the $Z$ basis. 
Charlie chooses $x=0,1$ to measure his qubit in a basis such that Charlie and Bob violate Bell locality, for example by choosing $-(X+Z)/\sqrt{2}$- and $(-Z+X)/\sqrt{2}$-measurements, such that the empirical probabilities $p(c_x,b_y|x,y)$ violate Bell locality~\cite{Bell_1976,brunner_bell_2014} and thus by Fine's theorem~\cite{abramsky2011sheaf,fine1982hidden,fine1982joint} a global distribution $p(c_0,c_1,b_0,b_1)$ reproducing the empirical correlations $p(c_x,b_y|x,y)$ cannot exist. 
Now, quantum theory predicts that Alice and Bob will always have the same outcome $a=b$ when $y=0$, as $\langle b, \lnot b |_{AB} (\ket{00}+\ket{11}) = 0$ for $b=0,1$. 
Therefore, we have that $p(a,b,c|x,y=0) = \delta_{a,b}p(b,c|x,y=0)$ and thus also \begin{equation} \label{eq:App_C_prob_cb_y0_is_prob_ca_y1}
\begin{split}
    p(c,b|x,y=0) &= \sum_{a} \delta_{a,b} p(c,b=a|x,y=0) \\ &= \sum_a \delta_{a,b} p(c,a|x,y=0) \\ &= \sum_a \delta_{a,b} p(c,a|x,y=1),     
\end{split}
\end{equation} where we used no-signalling to conclude that $p(c,a|x,y) = p(c,a|x)$; we can let Alice and Charlie perform their measurement before Bob's choice so that they obtain $a,c$ spacelike separated from Bob's choice $y$. 
This implies that the empirical correlations of $p(c,b|x,y=0)$ are those of $\delta_{a,b} p(c,a|x,y=1)$. Therefore, the joint correlation $p(c,b|x,y=0)$ is reproduced by $p(c,a|x,y=1)$ under $a \leftrightarrow b$ so that 
both $p(c,a|x,y=1),p(c,b|x,y=1)$  are marginals of the global distribution $p(a,b,c|x,y=1)$, which is empirically accessible if Alice, Bob and Charlie can meet inside the black hole.
If such a distribution $p(a,b,c|x,y=1)$ exists, then by a gluing procedure~\cite{walleghem2024connecting,fine1982joint,haddara2024local} we can produce a global distribution reproducing all empirical correlations $p(c,a|x=1,y=1),p(c,a|x=0,y=1),p(c,b|x=1,y=1),p(c,b|x=0,y=1)$, which by \cref{eq:App_C_prob_cb_y0_is_prob_ca_y1} are the same empirical correlations as the empirical Bell scenario correlations $p(c,b|x,y)$ for $x,y \in \{0,1\}$ that violate Bell locality with $c,a|x,y=1$ replaced by $c,b|x,y=0$.
However, as we explained a few lines ago, by Fine's theorem such global distribution cannot exist, and we find a contradiction. 

The gluing procedure unfolds as follows. Denoting $c|x$ by $c_x$, for example using $c_0$ for the outcome variable $c$ when $x=0$, the correlations $p(c_0,b,a|x=0,y=1)$ can be written as \begin{equation}
   \begin{split}
       p(c_0,&b,a|x{=}0,y{=}1) = p(a,b|x{=}0,y{=}1)p(c_0|x{=}0,y{=}1,b,a) \\ &= p(a,b|x{=}1,y{=}1)p(c_0|x{=}0,y{=}1,b,a),
   \end{split} 
\end{equation} using no-signalling for $p(a,b|x=0,y=1)=p(a,b|x=1,y=1)$, and noting that $p(c_0,b,a|x=0,y=1)$ is also empirically accessible if an observer as Charlie can obtain all of $c_0,b,a$ inside the black hole. 
We also have \begin{equation}
   \begin{split}
       p(c_1,b,a|x{=}1,y{=}1) = p(a,b|x{=}1,y{=}1)p(c_1|x{=}1,y{=}1,b,a),
   \end{split} 
\end{equation}
Therefore, the empirical correlations $p(c_0,a|x=0,y=1),p(c_0,b|x=0,y=1),p(c_1,a|x=1,y=1),p(c_1,b|x=1,y=1)$ can all be obtained from the ``glued'' global distribution \begin{equation}
    p(c_0,c_1,a,b) := p(c_1,a,b|x{=}1,y{=}1) p(c_0|a,b,x{=}0,y{=}1).
\end{equation}

We thus obtain the following result.

\begin{theorem}[No-signalling firewall argument]
    Assuming (Fi)-(Fiii) of \Cref{th:firewall}, an operational violation of no-signalling is found if Alice, Bob and Charlie can meet inside the black hole.
\end{theorem}

The meaning of no-signalling is determined by the semiclassical causal structure.
By an `operational' violation of no-signalling it is meant that this violation is experimental, witnessed by a single observer, i.e. that all relevant variables are jointly accessible (by Alice, Bob and Charlie if they can meet up inside the black hole). 
If Alice, Bob and Charlie cannot meet up inside the black hole, one could argue the violation is one of Local Agency~\cite{cavalcanti2021implications} rather than no-signalling.

\bibliography{refs}

@article{Hayward_2006,
   title={Formation and Evaporation of Nonsingular Black Holes},
   volume={96},
   ISSN={1079-7114},
   url={http://dx.doi.org/10.1103/PhysRevLett.96.031103},
   DOI={10.1103/physrevlett.96.031103},
   pages={031103},
   journal={Phys. Rev. Lett.},
   publisher={American Physical Society (APS)},
   author = {S. A. Hayward},
   year={2006},
   month=jan }

@article{Laddha_2021,
   title={The Holographic Nature of Null Infinity},
   volume={10},
   ISSN={2542-4653},
   url={http://dx.doi.org/10.21468/SciPostPhys.10.2.041},
   DOI={10.21468/scipostphys.10.2.041},
   pages={041},
   journal={Scipost Phys.},
   publisher={Stichting SciPost},
   author = {A. Laddha and S. Prabhu and S. Raju and P. Shrivastava},
   year={2021},
   month=feb }

@misc{goeller2022diffeomorphisminvariantobservablesdynamicalframes,
      title={Diffeomorphism-invariant observables and dynamical frames in gravity: reconciling bulk locality with general covariance}, 
      author = {C. Goeller and P. A. Höhn and J. Kirklin},
      year={2022},
      eprint={2206.01193},
      archivePrefix={arXiv},
      primaryClass={hep-th}}

@article{legget20005measurement,
author = {A. J. Leggett},
title = {The Quantum Measurement Problem},
journal = {Science},
volume = {307},
number = {5711},
pages = {871-872},
year = {2005},
doi = {10.1126/science.1109541},
URL = {https://www.science.org/doi/abs/10.1126/science.1109541}}

@article{Hance_2022,
   title={What does it take to solve the measurement problem?},
   volume={6},
   ISSN={2399-6528},
   url={http://dx.doi.org/10.1088/2399-6528/ac96cf},
   DOI={10.1088/2399-6528/ac96cf},
   number={10},
   journal={J. Phys. Commun.},
   publisher={IOP Publishing},
   author = {J. R. Hance and S. Hossenfelder},
   year={2022},
   month=oct, pages={102001} }

@misc{ormrod2024quantum,
      title={Quantum influences and event relativity}, 
      author = {N. Ormrod and J. Barrett},
      year={2024},
      eprint={2401.18005},
      archivePrefix={arXiv},
      primaryClass={quant-ph}}

@article{Maudlin1995,
	author = {T. Maudlin},
	doi = {10.1007/bf00763473},
	journal = {Topoi},
	number = {1},
	pages = {7--15},
	title = {Three Measurement Problems},
	volume = {14},
	year = {1995}
}

@article{hiscock1981parttwo,
  title = {Models of evaporating black holes. {II.} {E}ffects of the outgoing created radiation},
  author = {W. A. Hiscock},
  journal = {Phys. Rev. D},
  volume = {23},
  issue = {12},
  pages = {2823},
  numpages = {0},
  year = {1981},
  month = {Jun},
  publisher = {American Physical Society},
  doi = {10.1103/PhysRevD.23.2823},
  url = {https://link.aps.org/doi/10.1103/PhysRevD.23.2823}
}

@article{Barbado_2016,
   title={Hawking versus {U}nruh effects, or the difficulty of slowly crossing a black hole horizon},
   volume={2016},
   ISSN={1029-8479},
   url={http://dx.doi.org/10.1007/JHEP10(2016)161},
   DOI={10.1007/jhep10(2016)161},
   number={10},
   journal={J. High Energy Phys.},
   publisher={Springer Science and Business Media LLC},
   author = {L. C. Barbado and C. Barceló and L. J. Garay and G. Jannes},
   year={2016},
   month=oct }

@misc{maibach2026balancefluxlawsgeneral,
      title={Balance flux laws beyond general relativity}, 
      author={David Maibach and Jann Zosso},
      year={2026},
      eprint={2601.07091},
      archivePrefix={arXiv},
      primaryClass={gr-qc}}

@article{thorne1992gravitational,
  title = {Gravitational-wave bursts with memory: The Christodoulou effect},
  author = {Thorne, Kip S.},
  journal = {Phys. Rev. D},
  volume = {45},
  issue = {2},
  pages = {520--524},
  numpages = {0},
  year = {1992},
  month = {Jan},
  publisher = {American Physical Society},
  doi = {10.1103/PhysRevD.45.520},
  url = {https://link.aps.org/doi/10.1103/PhysRevD.45.520}
}

@article{kabel2024identification,
   title={Quantum coordinates, localisation of events, and the quantum hole argument},
   volume={8},
   ISSN={2399-3650},
   url={http://dx.doi.org/10.1038/s42005-025-02084-3},
   DOI={10.1038/s42005-025-02084-3},
   pages={185},
   journal={Commun. Phys.},
   publisher={Springer Science and Business Media LLC},
   author = {V. Kabel and A. de la Hamette and L. Apadula and C. Cepollaro and H. Gomes and J. Butterfield and {\v{C}}. Brukner},
   year={2025},
   month=apr }

@article{Ashtekar_2011,
  title = {Evaporation of two-dimensional black holes},
  author = {A. Ashtekar and F. Pretorius and Ramazanoğlu, F. M.},
  journal = {Phys. Rev. D},
  volume = {83},
  issue = {4},
  pages = {044040},
  numpages = {18},
  year = {2011},
  month = {Feb},
  publisher = {American Physical Society},
  doi = {10.1103/PhysRevD.83.044040},
  url = {https://link.aps.org/doi/10.1103/PhysRevD.83.044040}
}

@article{hiscock1981models,
  title={Models of evaporating black holes. {I}},
  author = {W. A. Hiscock},
  journal={Phys. Rev. D},
  volume={23},
  number={12},
  pages={2813},
  year={1981},
  publisher={APS},
  doi={10.1103/PhysRevD.23.2813}
}

@article{foo2022quantum,
  title={Quantum signatures of black hole mass superpositions},
  author = {J. Foo and C. S. Arabaci and M. Zych and R. B. Mann},
  journal={Phys. Rev. Lett.},
  volume={129},
  number={18},
  pages={181301},
  year={2022},
  publisher={APS},
 url ={https://doi.org/10.1103/PhysRevLett.129.181301}
}

@article{henderson2018harvesting,
  title={Harvesting entanglement from the black hole vacuum},
  author = {L. J. Henderson and R. A. Hennigar and R. B. Mann and A. R. Smith and J. Zhang},
  journal={Class. Quantum Gravity},
  volume={35},
  number={21},
  pages={21LT02},
  year={2018},
  publisher={IOP Publishing},
  doi={10.1088/1361-6382/aae27e}
}

@article{Stritzelberger_2021,
   title = {Entanglement harvesting with coherently delocalized matter},
  author = {N. Stritzelberger and L. J. Henderson and V. Baccetti and N. C. Menicucci and A. Kempf},
  journal = {Phys. Rev. D},
  volume = {103},
  issue = {1},
  pages = {016007},
  numpages = {14},
  year = {2021},
  month = {Jan},
  publisher = {American Physical Society},
  doi = {10.1103/PhysRevD.103.016007},
  url = {https://link.aps.org/doi/10.1103/PhysRevD.103.016007}
}

@article{Hotta_2020,
    title = {Duality in the dynamics of {U}nruh-{D}e{W}itt detectors in conformally related spacetimes},
  author = {M. Hotta and A. Kempf and Mart\'{\i}n-Mart\'{\i}nez, E. and Tomitsuka, T. and Yamaguchi, K.},
  journal = {Phys. Rev. D},
  volume = {101},
  issue = {8},
  pages = {085017},
  numpages = {10},
  year = {2020},
  month = {Apr},
  publisher = {American Physical Society},
  doi = {10.1103/PhysRevD.101.085017},
  url = {https://link.aps.org/doi/10.1103/PhysRevD.101.085017}
}

@article{Ju_rez_Aubry_2014,
   title={Onset and decay of the 1 + 1 {H}awking-{U}nruh effect: what the derivative-coupling detector saw},
   volume={31},
   ISSN={1361-6382},
   url={http://dx.doi.org/10.1088/0264-9381/31/24/245007},
   DOI={10.1088/0264-9381/31/24/245007},
   number={24},
   journal={Class. Quantum Gravity},
   publisher={IOP Publishing},
   author = {B. A. Juárez-Aubry and J. Louko},
   year={2014},
   month=nov, pages={245007} }

@article{Stritzelberger_2020,
   title = {Coherent delocalization in the light-matter interaction},
  author = {N. Stritzelberger and A. Kempf},
  journal = {Phys. Rev. D},
  volume = {101},
  issue = {3},
  pages = {036007},
  numpages = {10},
  year = {2020},
  month = {Feb},
  publisher = {American Physical Society},
  doi = {10.1103/PhysRevD.101.036007},
  url = {https://link.aps.org/doi/10.1103/PhysRevD.101.036007}
}

@article{Ng_2014,
   title={{U}nruh-{D}e{W}itt detector response along static and circular-geodesic trajectories for {S}chwarzschild–anti-de {S}itter black holes},
  author = {K. K. Ng and L. Hodgkinson and J. Louko and R. B. Mann and Mart\'{\i}n-Mart\'{\i}nez, Eduardo},
  journal = {Phys. Rev. D},
  volume = {90},
  issue = {6},
  pages = {064003},
  numpages = {13},
  year = {2014},
  month = {Sep},
  publisher = {American Physical Society},
  doi = {10.1103/PhysRevD.90.064003},
  url = {https://link.aps.org/doi/10.1103/PhysRevD.90.064003}
}

@article{Henderson_2020,
   title={Anti-{H}awking phenomena},
   volume={809},
   ISSN={0370-2693},
   url={http://dx.doi.org/10.1016/j.physletb.2020.135732},
   DOI={10.1016/j.physletb.2020.135732},
   journal={Phys. Lett. B},
   publisher={Elsevier BV},
   author = {L. J. Henderson and R. A. Hennigar and R. B. Mann and A. R. Smith and J. Zhang},
   year={2020},
   month=oct, pages={135732} }

@article{Cong_2020,
   title={Effects of horizons on entanglement harvesting},
   volume={2020},
   ISSN={1029-8479},
   url={http://dx.doi.org/10.1007/JHEP10(2020)067},
   DOI={10.1007/jhep10(2020)067},
   number={},
    pages = {10},
   journal={J. High Energy Phys.},
   author = {W. Cong and C. Qian and M. R. Good and R. B. Mann},
   year={2020},
   month=oct }

@article{Ng_2018,
   title = {New techniques for entanglement harvesting in flat and curved spacetimes},
  author = {K. K. Ng and R. B. Mann and Mart\'{\i}n-Mart\'{\i}nez, Eduardo},
  journal = {Phys. Rev. D},
  volume = {97},
  issue = {12},
  pages = {125011},
  numpages = {8},
  year = {2018},
  month = {Jun},
  publisher = {American Physical Society},
  doi = {10.1103/PhysRevD.97.125011},
  url = {https://link.aps.org/doi/10.1103/PhysRevD.97.125011}
}

@article{vsoda2022acceleration,
  title={Acceleration-induced effects in stimulated light-matter interactions},
  author = {{\v{S}}oda, Barbara and Sudhir, Vivishek and Kempf, Achim},
  journal={Phys. Rev. Lett.},
  volume={128},
  number={16},
  pages={163603},
  year={2022},
  publisher={APS},
  doi={10.1103/PhysRevLett.128.163603}
}

@article{pan2024gravity,
  title={Gravity-induced transparency},
  author = {Y. Pan and B. Zhang},
  journal={Phys. Rev.  D},
  volume={109},
  number={12},
  pages={125018},
  year={2024},
  publisher={APS}, 
  doi={10.1103/PhysRevD.109.125018}
}

@article{Henderson_2022,
    author = {L. J. Henderson and S. Y. Ding and R. B. Mann},
    title = {Entanglement harvesting with a twist},
    journal = {AVS Quantum Science},
    volume = {4},
    number = {1},
    pages = {014402},
    year = {2022},
    month = {02},
    issn = {2639-0213},
    doi = {10.1116/5.0078314},
    url = {https://doi.org/10.1116/5.0078314}
}

@article{Gallock_Yoshimura_2021,
   title = {Harvesting entanglement with detectors freely falling into a black hole},
  author = {K. Gallock-Yoshimura and E. Tjoa and R. B. Mann},
  journal = {Phys. Rev. D},
  volume = {104},
  issue = {2},
  pages = {025001},
  numpages = {19},
  year = {2021},
  month = {Jul},
  publisher = {American Physical Society},
  doi = {10.1103/PhysRevD.104.025001},
  url = {https://link.aps.org/doi/10.1103/PhysRevD.104.025001}
}

@article{Mendez_Avalos_2022,
   title={Entanglement harvesting of three {U}nruh-{D}e{W}itt detectors},
   volume={54},
    pages = {87},
   ISSN={1572-9532},
   url={http://dx.doi.org/10.1007/s10714-022-02956-x},
   DOI={10.1007/s10714-022-02956-x},
   number={8},
   journal={Gen. Relativ. Gravit.},
   publisher={Springer Science and Business Media LLC},
   author = {D. Mendez-Avalos and L. J. Henderson and K. Gallock-Yoshimura and R. B. Mann},
   year={2022},
   month=aug }

@article{Smith_2014,
   title={Looking inside a black hole},
   volume={31},
   ISSN={1361-6382},
   url={http://dx.doi.org/10.1088/0264-9381/31/8/082001},
   DOI={10.1088/0264-9381/31/8/082001},
   number={8},
   journal={Class. Quantum Gravity},
   publisher={IOP Publishing},
   author = {A. R. H. Smith and R. B. Mann},
   year={2014},
   month=apr, pages={082001} }

@article{Tjoa_2020,
   title={Harvesting correlations in {S}chwarzschild and collapsing shell spacetimes},
   volume={2020},
   ISSN={1029-8479},
   url={http://dx.doi.org/10.1007/JHEP08(2020)155},
   DOI={10.1007/jhep08(2020)155},
   number={8},
    pages = {155},
   journal={J. High Energy Phys.},
   author = {E. Tjoa and R. B. Mann},
   year={2020},
   month=aug }

@article{Hodgkinson_2012,
   title={Static, stationary, and inertial {U}nruh-{D}e{W}itt detectors on the {BTZ} black hole},
  author = {L. Hodgkinson and J. Louko},
  journal = {Phys. Rev. D},
  volume = {86},
  issue = {6},
  pages = {064031},
  numpages = {16},
  year = {2012},
  month = {Sep},
  publisher = {American Physical Society},
  doi = {10.1103/PhysRevD.86.064031},
  url = {https://link.aps.org/doi/10.1103/PhysRevD.86.064031}
}

@misc{robbins2020entanglementamplificationrotatingblack,
      title={Entanglement Amplification from Rotating Black Holes}, 
      author = {M. P. G. Robbins and L. J. Henderson and R. B. Mann},
      year={2020},
      eprint={2010.14517},
      archivePrefix={arXiv},
      primaryClass={hep-th}
}

@misc{niermann2024particle,
  title={Particle detectors in superposition in de {S}itter spacetime},
  author = {L. Niermann and L. C. Barbado},
  eprint={2403.02087},
archivePrefix={arXiv},
year = {2024},
      primaryClass={gr-qc}
}

@article{juarez2022quantumkicks,
    author = {B. A. Juárez-Aubry and J. Louko},
    title = {Quantum kicks near a {C}auchy horizon},
    journal = {AVS Quantum Science},
    volume = {4},
    number = {1},
    pages = {013201},
    year = {2022},
    month = {02},
    issn = {2639-0213},
    doi = {10.1116/5.0073373},
    url = {https://doi.org/10.1116/5.0073373}
}

@article{Hollands_2020,
   title={Quantum instability of the {C}auchy horizon in {R}eissner–{N}ordström–de{S}itter spacetime},
   volume={37},
   ISSN={1361-6382},
   url={http://dx.doi.org/10.1088/1361-6382/ab8052},
   DOI={10.1088/1361-6382/ab8052},
   number={11},
   journal={Class. Quantum Gravity},
   publisher={IOP Publishing},
   author = {S. Hollands and R. M. Wald and J. Zahn},
   year={2020},
   month=may, pages={115009} }

@article{hollands2020quantum,
  title = {Quantum stress tensor at the Cauchy horizon of the Reissner--Nordstr\"om--de {S}itter spacetime},
  author = {S. Hollands and C. Klein and J. Zahn},
  journal = {Phys. Rev. D},
  volume = {102},
  issue = {8},
  pages = {085004},
  numpages = {10},
  year = {2020},
  month = {Oct},
  publisher = {American Physical Society},
  doi = {10.1103/PhysRevD.102.085004},
  url = {https://link.aps.org/doi/10.1103/PhysRevD.102.085004}
}

@article{cardoso2018quasi,
  title = {Quasinormal Modes and Strong Cosmic Censorship},
  author = {V. Cardoso and J. L. Costa and K. Destounis and P. Hintz and A. Jansen},
  journal = {Phys. Rev. Lett.},
  volume = {120},
  issue = {3},
  pages = {031103},
  numpages = {6},
  year = {2018},
  month = {Jan},
  publisher = {American Physical Society},
  doi = {10.1103/PhysRevLett.120.031103},
  url = {https://link.aps.org/doi/10.1103/PhysRevLett.120.031103}
}

@article{Dias_2019strong,
   title={Strong cosmic censorship for charged de {S}itter black holes with a charged scalar field},
   volume={36},
   ISSN={1361-6382},
   url={http://dx.doi.org/10.1088/1361-6382/aafcf2},
   DOI={10.1088/1361-6382/aafcf2},
   number={4},
   journal={Class. Quantum Gravity},
   publisher={IOP Publishing},
   author = {O. J. C. Dias and H. S. Reall and J. E. Santos},
   year={2019},
   month=jan, pages={045005} }

@article{Emparan_2020,
   title={Strong cosmic censorship in the {BTZ} black hole},
   volume={2020},
   ISSN={1029-8479},
   url={http://dx.doi.org/10.1007/JHEP06(2020)038},
   DOI={10.1007/jhep06(2020)038},
   number={6},
   journal={J. High Energy Phys.},
   publisher={Springer Science and Business Media LLC},
   author = {R. Emparan and M. Tomašević},
   year={2020},
   month=jun }

@article{mcmaken2023,
  title = {Semiclassical instability of inner-extremal regular black holes},
  author = {T. McMaken},
  journal = {Phys. Rev. D},
  volume = {107},
  issue = {12},
  pages = {125023},
  numpages = {16},
  year = {2023},
  month = {Jun},
  publisher = {American Physical Society},
  doi = {10.1103/PhysRevD.107.125023},
  url = {https://link.aps.org/doi/10.1103/PhysRevD.107.125023}
}

@article{Dias_2019,
   title={The {BTZ} black hole violates strong cosmic censorship},
   volume={2019},
   ISSN={1029-8479},
   url={http://dx.doi.org/10.1007/JHEP12(2019)097},
   DOI={10.1007/jhep12(2019)097},
   number={12},
   journal={J. High Energy Phys.},
   publisher={Springer Science and Business Media LLC},
   author = {O. J. Dias and H. S. Reall and J. E. Santos},
   year={2019},
   month=dec }

@article{Papadodimas_2020,
   title={A simple quantum test for smooth horizons},
   volume={2020},
   ISSN={1029-8479},
   url={http://dx.doi.org/10.1007/JHEP12(2020)003},
   DOI={10.1007/jhep12(2020)003},
   number={12},
   journal={J. High Energy Phys.},
   publisher={Springer Science and Business Media LLC},
   author = {K. Papadodimas and S. Raju and P. Shrivastava},
   year={2020},
   month=dec }

@article{Lanir_2019,
  title = {Analysis of quantum effects inside spherical charged black holes},
  author = {A. Lanir and A. Ori and N. Zilberman and O. Sela and A. Maline and A. Levi},
  journal = {Phys. Rev. D},
  volume = {99},
  issue = {6},
  pages = {061502},
  numpages = {6},
  year = {2019},
  month = {Mar},
  publisher = {American Physical Society},
  doi = {10.1103/PhysRevD.99.061502},
  url = {https://link.aps.org/doi/10.1103/PhysRevD.99.061502}
}

@article{Ju_rez_Aubry_2015,
   title={Can a particle detector cross a Cauchy horizon?},
   volume={24},
   ISSN={1793-6594},
   url={http://dx.doi.org/10.1142/S0218271815420055},
   DOI={10.1142/s0218271815420055},
   number={09},
   journal={Int. J. Modern Phys. D},
   publisher={World Scientific Pub Co Pte Lt},
   author = {B. A. Juárez-Aubry},
   year={2015},
   month=jul, pages={1542005} }

@article{zilberman2020quantumfluxes,
  title = {Quantum Fluxes at the Inner Horizon of a Spherical Charged Black Hole},
  author = {N. Zilberman and A. Levi and A. Ori},
  journal = {Phys. Rev. Lett.},
  volume = {124},
  issue = {17},
  pages = {171302},
  numpages = {7},
  year = {2020},
  month = {Apr},
  publisher = {American Physical Society},
  doi = {10.1103/PhysRevLett.124.171302},
  url = {https://link.aps.org/doi/10.1103/PhysRevLett.124.171302}
}

@article{Goto_2021,
   title={Replica wormholes for an evaporating 2D black hole},
   volume={2021},
   ISSN={1029-8479},
   url={http://dx.doi.org/10.1007/JHEP04(2021)289},
   DOI={10.1007/jhep04(2021)289},
   number={4},
   journal={J. High Energy Phys.},
   publisher={Springer Science and Business Media LLC},
   author = {K. Goto and T. Hartman and A. Tajdini},
   year={2021},
   month=apr }

@article{Hartman_2020,
   title={Islands in asymptotically flat 2D gravity},
   volume={2020},
   ISSN={1029-8479},
   url={http://dx.doi.org/10.1007/JHEP07(2020)022},
   DOI={10.1007/jhep07(2020)022},
   number={7},
   journal={J. High Energy Phys.},
   publisher={Springer Science and Business Media LLC},
   author = {T. Hartman and E. Shaghoulian and A. Strominger},
   year={2020},
   month=jul }

@article{wang2021page,
  title = {{P}age curves for a family of exactly solvable evaporating black holes},
  author = {X. Wang and R. Li and J. Wang},
  journal = {Phys. Rev. D},
  volume = {103},
  issue = {12},
  pages = {126026},
  numpages = {17},
  year = {2021},
  month = {Jun},
  publisher = {American Physical Society},
  doi = {10.1103/PhysRevD.103.126026},
  url = {https://link.aps.org/doi/10.1103/PhysRevD.103.126026}
}

@article{Geng_2020,
   title={Massive islands},
   volume={2020},
   ISSN={1029-8479},
   url={http://dx.doi.org/10.1007/JHEP09(2020)121},
   DOI={10.1007/jhep09(2020)121},
   number={9},
   journal={J. High Energy Phys.},
   publisher={Springer Science and Business Media LLC},
   author = {H. Geng and A. Karch},
   year={2020},
   month=sep }

@article{kibe2022holographic,
  title={Holographic spacetime, black holes and quantum error correcting codes: a review},
  author = {T. Kibe and P. Mandayam and A. Mukhopadhyay},
  journal={Eur. Phys. J. C},
  volume={82},
  number={5},
  pages={463},
   DOI={10.1140/epjc/s10052-022-10382-1},
  year={2022},
  publisher={Springer}
}

@misc{chen2020quantumextremalislandseasyI,
      title={Quantum Extremal Islands Made Easy, Part {I}: Entanglement on the Brane}, 
      author = {H. Z. Chen and R. C. Myers and D. Neuenfeld and I. A. Reyes and J. Sandor},
      year={2020},
      eprint={2006.04851},
      archivePrefix={arXiv},
      primaryClass={hep-th}}

@misc{almheiri2023islandsoutsidehorizon,
      title={Islands outside the horizon}, 
      author = {A. Almheiri and R. Mahajan and J. Maldacena},
      year={2019},
      eprint={1910.11077},
      archivePrefix={arXiv},
      primaryClass={hep-th}}

@misc{chen2020quantumextremalislandseasyII,
      title={Quantum Extremal Islands Made Easy, Part {II}: Black Holes on the Brane}, 
      author = {H. Z. Chen and R. C. Myers and D. Neuenfeld and I. A. Reyes and J. Sandor},
      year={2020},
      eprint={2010.00018},
      archivePrefix={arXiv},
      primaryClass={hep-th}}

@Article{almheiri2020entanglementislandsinhigher,
	title={Entanglement islands in higher dimensions},
	author = {A. Almheiri and R. Mahajan and J. E. Santos},
	journal={SciPost Phys.},
	volume={9},
	pages={001},
	year={2020},
	publisher={SciPost},
	doi={10.21468/SciPostPhys.9.1.001},
	url={https://scipost.org/10.21468/SciPostPhys.9.1.001},
}

@article{Hashimoto_2020,
   title={Islands in {S}chwarzschild black holes},
   volume={2020},
   ISSN={1029-8479},
   url={http://dx.doi.org/10.1007/JHEP06(2020)085},
   DOI={10.1007/jhep06(2020)085},
   number={6},
   journal={J. High Energy Phys.},
   publisher={Springer Science and Business Media LLC},
   author = {K. Hashimoto and N. Iizuka and Y. Matsuo},
   year={2020},
   month=jun }

@misc{geng2025revisitingrecentprogresskarchrandall,
      title={Revisiting Recent Progress in the {K}arch--{R}andall Braneworld}, 
      author = {H. Geng},
      year={2025},
      eprint={2306.15671},
      archivePrefix={arXiv},
      primaryClass={hep-th},
}

@article{wang2024entanglement,
  title = {Entanglement islands and the {P}age curve of {H}awking radiation for rotating {K}err black holes},
  author = {L. Wang and R. Li},
  journal = {Phys. Rev. D},
  volume = {110},
  issue = {6},
  pages = {066012},
  numpages = {15},
  year = {2024},
  month = {Sep},
  publisher = {American Physical Society},
  doi = {10.1103/PhysRevD.110.066012},
  url = {https://link.aps.org/doi/10.1103/PhysRevD.110.066012}
}

@misc{akers2022blackholeinteriornonisometric,
      title={The black hole interior from non-isometric codes and complexity}, 
      author = {C. Akers and N. Engelhardt and D. Harlow and G. Penington and S. Vardhan},
      year={2022},
      eprint={2207.06536},
      archivePrefix={arXiv},
      primaryClass={hep-th}}

@misc{engelhardt2023algebraicereprcomplexitytransfer,
      title={Algebraic {ER}={EPR} and {C}omplexity {T}ransfer}, 
      author = {N. Engelhardt and H. Liu},
      year={2023},
      eprint={2311.04281},
      archivePrefix={arXiv},
      primaryClass={hep-th}}

@article{zilberman2022quantum,
  title = {Quantum Fluxes at the Inner Horizon of a Spinning Black Hole},
  author = {N. Zilberman and M. Casals and A. Ori and A. C. Ottewill},
  journal = {Phys. Rev. Lett.},
  volume = {129},
  issue = {26},
  pages = {261102},
  numpages = {6},
  year = {2022},
  month = {Dec},
  publisher = {American Physical Society},
  doi = {10.1103/PhysRevLett.129.261102},
  url = {https://link.aps.org/doi/10.1103/PhysRevLett.129.261102}
}

@article{wang2024singular,
  title={Singular excitement beyond the horizon of a rotating black hole},
  author = {S. Wang and M. R. Preciado-Rivas and Spadafora, Massimiliano and Mann, Robert B},
  journal={Phys. Rev.  D},
  volume={110},
  number={6},
  pages={065013},
  year={2024},
  publisher={APS},
  doi={10.1103/PhysRevD.110.065013}
}

@article{paczos2023hawking,
  title={{H}awking radiation for detectors in superposition of locations outside a black hole},
  author = {J. Paczos and L. C. Barbado},
  journal={Phys. Rev.  D},
  volume={108},
  number={12},
  pages={125015},
  year={2023},
  publisher={APS}, 
  doi={10.1103/PhysRevD.108.125015}
}

@article{Foo_2020,
   title={{U}nruh-{D}e{W}itt detectors in quantum superpositions of trajectories},
  author = {J. Foo and S. Onoe and M. Zych},
  journal = {Phys. Rev. D},
  volume = {102},
  issue = {8},
  pages = {085013},
  numpages = {15},
  year = {2020},
  month = {Oct},
  publisher = {American Physical Society},
  doi = {10.1103/PhysRevD.102.085013},
  url = {https://link.aps.org/doi/10.1103/PhysRevD.102.085013}
}

@misc{suryaatmadja2023signaturesrotatingblackholes,
      title={Signatures of Rotating Black Holes in Quantum Superposition}, 
      author = {C. Suryaatmadja and C. S. Arabaci and M. P. G. Robbins and J. Foo and M. Zych and R. B. Mann},
      year={2023},
      eprint={2310.10864},
      archivePrefix={arXiv},
      primaryClass={gr-qc}
}

@misc{chen2024quantum,
  title={Quantum effects in gravity beyond the Newton potential from a delocalised quantum source},
  author = {L. Chen and F. Giacomini},
      year={2024},
      eprint={2402.10288},
      archivePrefix={arXiv},
      primaryClass={quant-ph}
}

@article{chen2023quantum,
  title={Quantum states of fields for quantum split sources},
  author = {L. Chen and F. Giacomini and C. Rovelli},
  journal={Quantum},
  volume={7},
  pages={958},
  year={2023},
  publisher={Verein zur F{\"o}rderung des Open Access Publizierens in den Quantenwissenschaften},
  doi={10.22331/q-2023-03-20-958}
}

@article{spadafora2024deepknottedblackhole,
      title={Deep in the Knotted Black Hole}, 
      author = {M. Spadafora and M. Naeem and M. R. Preciado-Rivas and R. B. Mann and J. Louko},
        journal = {Phys. Rev. D},
  volume = {111},
  issue = {6},
  pages = {065013},
  numpages = {16},
  year = {2025},
  month = {Mar},
  publisher = {American Physical Society},
  doi = {10.1103/PhysRevD.111.065013}
}

@article{Witten_2018,
   title={APS Medal for Exceptional Achievement in Research: Invited article on entanglement properties of quantum field theory},
   volume={90},
   ISSN={1539-0756},
   url={http://dx.doi.org/10.1103/RevModPhys.90.045003},
   DOI={10.1103/revmodphys.90.045003},
   number={4},
   journal={Reviews of Modern Physics},
   publisher={American Physical Society (APS)},
   author={Witten, Edward},
   year={2018},
   month=Oct }

@book{Hollands_2018,
   title={Entanglement Measures and Their Properties in Quantum Field Theory},
   ISBN={9783319949024},
   ISSN={2197-1765},
   url={http://dx.doi.org/10.1007/978-3-319-94902-4},
   DOI={10.1007/978-3-319-94902-4},
   journal={SpringerBriefs in Mathematical Physics},
   publisher={Springer International Publishing},
   author={Hollands, Stefan and Sanders, Ko},
   year={2018} }

@article{braunstein2013better,
  title={Better late than never: information retrieval from black holes},
  author = {S. L. Braunstein and S. Pirandola and {\.Z}yczkowski, Karol},
  journal={Phys. Rev. Lett.},
  volume={110},
  number={10},
  pages={101301},
  year={2013},
  publisher={APS},
  doi={10.1103/PhysRevLett.110.101301}
}

@article{braunstein2007quantum,
  title={Quantum information cannot be completely hidden in correlations: implications for the black-hole information paradox},
  author = {S. L. Braunstein and A. K. Pati},
  journal={Phys. Rev. Lett.},
  volume={98},
  number={8},
  pages={080502},
  year={2007},
  publisher={APS},
  doi={10.1103/PhysRevLett.98.080502}
}

@article{hawking1975particle,
  title={Particle creation by black holes},
  author = {S. W. Hawking},
  journal={Commun. Math. Phys.},
  volume={43},
  number={3},
  pages={199--220},
  year={1975},
  publisher={Springer},
 doi={10.1007/BF02345020}
}

@article{calmet2022brief,
  title={A brief history of {H}awking’s information paradox},
  author = {X. Calmet and S. D. Hsu},
  journal={Europhys. Lett.},
  year={2022},
    volume={139},
pages={49001},
  publisher={IOP Publishing},
  doi={10.1209/0295-5075/ac81e8}
}

@article{mathur2009information,
  title={The information paradox: a pedagogical introduction},
  author = {S. D. Mathur},
  journal={Class. Quantum Gravity},
  volume={26},
  number={22},
  pages={224001},
  year={2009},
  publisher={IOP Publishing},
  doi={10.1088/0264-9381/26/22/224001}
}

@article{giddings2006black,
  title={Black hole information, unitarity, and nonlocality},
  author = {S. B. Giddings},
  journal={Phys. Rev.  D},
  volume={74},
  number={10},
  pages={106005},
  year={2006},
  publisher={APS}, 
  doi={10.1103/PhysRevD.74.106005}
}

@article{giddings2013nonviolent,
  title={Nonviolent information transfer from black holes: A field theory parametrization},
  author = {S. B. Giddings},
  journal={Phys. Rev.  D},
  volume={88},
  number={2},
  pages={024018},
  year={2013},
  publisher={APS},
  doi={10.1103/PhysRevD.88.024018}
}

@article{susskind2016computational,
  title={Computational complexity and black hole horizons},
  author = {L. Susskind},
  journal={Fortschr. Phys.},
  volume={64},
  number={1},
  pages={24--43},
  year={2016},
  publisher={Wiley Online Library},
doi={10.1002/prop.201500092}
}

@article{almheiri2021entropy,
  title={The entropy of {H}awking radiation},
  author = {A. Almheiri and T. Hartman and J. Maldacena and E. Shaghoulian and A. Tajdini},
  journal={Rev. Modern Phys.},
  volume={93},
  number={3},
  pages={035002},
  year={2021},
  publisher={APS},
  doi={10.1103/RevModPhys.93.035002}
}

@article{penington2022replica,
  title={Replica wormholes and the black hole interior},
  author = {G. Penington and S. H. Shenker and D. Stanford and Z. Yang},
  journal={J. High Energy Phys.},
  volume={2022},
  number={3},
  pages={1--87},
  year={2022},
  publisher={Springer},
  doi={10.48550/arXiv.1911.11977}
}

@article{barcelo2005analogue,
   title={Analogue Gravity},
   volume={8},
   ISSN={1433-8351},
   url={http://dx.doi.org/10.12942/lrr-2005-12},
   DOI={10.12942/lrr-2005-12},
   number={1},
    volume  = {8},
  pages = {12},
   journal={Liv. Rev. Relativ.},
   publisher={Springer Science and Business Media LLC},
   author = {C. Barceló and S. Liberati and M. Visser},
   year={2005},
   month=dec }

@article{Barbado_2011,
   title={{H}awking radiation as perceived by different observers},
   volume={28},
   ISSN={1361-6382},
   url={http://dx.doi.org/10.1088/0264-9381/28/12/125021},
   DOI={10.1088/0264-9381/28/12/125021},
   number={12},
   journal={Class. Quantum Gravity},
   publisher={IOP Publishing},
   author = {L. C. Barbado and C. Barceló and L. J. Garay},
   year={2011},
   month=may, pages={125021} }

@article{preciado2024more,
     title = {More excitement across the horizon},
  author = {M. R. Preciado-Rivas and Naeem, M. and Mann, R. B. and Louko, J.},
  journal = {Phys. Rev. D},
  volume = {110},
  issue = {2},
  pages = {025002},
  numpages = {15},
  year = {2024},
  month = {Jul},
  publisher = {American Physical Society},
  doi = {10.1103/PhysRevD.110.025002},
  url = {https://link.aps.org/doi/10.1103/PhysRevD.110.025002}
}

@article{foo2023quantum,
  title={Quantum superpositions of Minkowski spacetime},
  author = {J. Foo and C. S. Arabaci and M. Zych and R. B. Mann},
  journal={Phys. Rev.  D},
  volume={107},
  number={4},
  pages={045014},
  year={2023},
  publisher={APS},
  doi={10.1103/PhysRevD.107.045014}
}

@article{goel2024accelerateddetectorsuperposedspacetime,
      title={Accelerated detector in a superposed spacetime}, 
      author = {L. Goel and E. A. Patterson and M. R. Preciado-Rivas and M. Torabian and R. B. Mann and N. Afshordi},
       journal = {Phys. Rev. D},
  volume = {111},
  issue = {2},
  pages = {025015},
  numpages = {20},
  year = {2025},
  month = {Jan},
  publisher = {American Physical Society},
  doi = {10.1103/PhysRevD.111.025015},
  url = {https://link.aps.org/doi/10.1103/PhysRevD.111.025015}
}

@article{calmet2022quantuma,
  title={Quantum hair and black hole information},
  author = {X. Calmet and S. D. Hsu},
  journal={Phys. Lett. B},
  volume={827},
  pages={136995},
  year={2022},
  publisher={Elsevier},
  DOI={10.1016/j.physletb.2022.136995}
}

@article{almheiri2013black,
  title={Black holes: complementarity or firewalls?},
  author = {A. Almheiri and D. Marolf and J. Polchinski and J. Sully},
  journal={J. High Energy Phys.},
  volume={2013},
  number={2},
  pages={1--20},
  year={2013},
  publisher={Springer},
  doi={10.1007/JHEP02%282013%29062}
}

@article{hawking1976breakdown,
  title={Breakdown of predictability in gravitational collapse},
  author = {S. W. Hawking},
  journal={Phys. Rev.  D},
  volume={14},
  number={10},
  pages={2460},
  year={1976},
  publisher={APS}, 
  doi = {10.1103/PhysRevD.14.2460},
}

@article{giacomini2019quantum,
  title={Quantum mechanics and the covariance of physical laws in quantum reference frames},
  author = {F. Giacomini and E. Castro-Ruiz and {\v{C}}. Brukner},
  journal={Nat. Commun.},
  volume={10},
  number={1},
  pages={494},
  year={2019},
  publisher={Nature Publishing Group UK London},
  doi={10.1038/s41467-018-08155-0}
}

@article{susskind1993stretched,
  title = {The stretched horizon and black hole complementarity},
  author = {L. Susskind and L. Thorlacius and J. Uglum},
  journal = {Phys. Rev. D},
  volume = {48},
  issue = {8},
  pages = {3743--3761},
  numpages = {0},
  year = {1993},
  month = {Oct},
  publisher = {American Physical Society},
  doi = {10.1103/PhysRevD.48.3743},
  url = {https://link.aps.org/doi/10.1103/PhysRevD.48.3743}
}

@article{reeh1961bemerkungen,
  title={Bemerkungen zur unit{\"a}r{\"a}quivalenz von lorentzinvarianten feldern},
  author = {H. Reeh and S. Schlieder},
  journal={Il Nuovo Cimento (1955-1965)},
  volume={22},
  pages={1051--1068},
  year={1961},
  publisher={Springer},
doi={10.1007/BF01645904}
}

@article{Stephens_1994,
   title={Black hole evaporation without information loss},
   volume={11},
   ISSN={1361-6382},
   url={http://dx.doi.org/10.1088/0264-9381/11/3/014},
   DOI={10.1088/0264-9381/11/3/014},
   number={3},
   journal={Class. Quantum Gravity},
   publisher={IOP Publishing},
   author = {C. R. Stephens and G. 't Hooft and B. F. Whiting},
   year={1994},
   month=mar, pages={621–647} }

@misc{devuyst2025gravitationalentropyobserverdependent,
      title={Gravitational entropy is observer-dependent}, 
      author = {J. D. Vuyst and S. Eccles and P. A. Hoehn and J. Kirklin},
      year={2025},
      eprint={2405.00114},
      archivePrefix={arXiv},
      primaryClass={hep-th}}

@article{harlow2016jerusalem,
  title={Jerusalem lectures on black holes and quantum information},
  author = {D. Harlow},
  journal={Rev. Modern Phys.},
  volume={88},
  number={1},
  pages={015002},
  year={2016},
  publisher={APS},
  doi={10.1103/RevModPhys.88.015002}
}

@article{geng2022inconsistency,
  title={Inconsistency of islands in theories with long-range gravity},
  author = {H. Geng and A. Karch and C. Perez-Pardavila and S. Raju and L. Randall and M. Riojas and S. Shashi},
  journal={J. High Energy Phys.},
  volume={2022},
  number={1},
  pages={1--46},
  year={2022},
  publisher={Springer},   DOI={10.1007/jhep01(2022)182}

}

@article{penington2020entanglement,
  title={Entanglement wedge reconstruction and the information paradox},
  author = {G. Penington},
  journal={J. High Energy Phys.},
  volume={2020},
  number={9},
  pages={1--84},
  year={2020},
  publisher={Springer},
  doi={10.48550/arXiv.1905.08255}
}

@article{geng2021information,
  title={{Information transfer with a gravitating bath}},
	author = {H. Geng and A. Karch and C. Perez-Pardavila and S. Raju and L. Randall and M. Riojas and S. Shashi},
	journal={SciPost Phys.},
	volume={10},
	pages={103},
	year={2021},
	publisher={SciPost},
	doi={10.21468/SciPostPhys.10.5.103},
	url={https://scipost.org/10.21468/SciPostPhys.10.5.103}
}

@article{maldacena1999large,
  title={The large-{N} limit of superconformal field theories and supergravity},
  author = {J. Maldacena},
  journal={Int. J. Theor. Phys.},
  volume={38},
  number={4},
  pages={1113--1133},
  year={1999},
  publisher={Springer},
  doi={10.1023/A%3A1026654312961}
}

@article{chowdhury2022sachdev,
  title={{S}achdev--{Y}e--{K}itaev  models and beyond: Window into non-Fermi liquids},
  author = {D. Chowdhury and A. Georges and O. Parcollet and S. Sachdev},
  journal={Rev. Modern Phys.},
  volume={94},
  number={3},
  pages={035004},
  year={2022},   DOI={10.1103/revmodphys.94.035004},
  publisher={APS}
}

@article{mertens2023solvable,
  title={Solvable models of quantum black holes: a review on Jackiw--Teitelboim gravity},
  author = {T. G. Mertens and G. J. Turiaci},
  journal={Living Rev. Relativ.},
  volume={26},
  number={1},
  pages={4},
  year={2023},
  publisher={Springer},
  doi={10.1007/s41114-023-00046-1}
}

@article{shenker2014black,
  title={Black holes and the butterfly effect},
  author = {S. H. Shenker and D. Stanford},
  journal={J. High Energy Phys.},
  number={3},
  pages={067},
  volume={2014},
  year={2014},
  DOI={10.1007/jhep03(2014)067}
}

@article{page1993information,
  title={Information in black hole radiation},
  author = {D. N. Page},
  journal={Phys. Rev. Lett.},
  volume={71},
  number={23},
  pages={3743},
  year={1993},
  publisher={APS},
  doi={10.1103/PhysRevLett.71.3743}
}

@article{page1993average,
  title={Average entropy of a subsystem},
  author = {D. N. Page},
  journal={Phys. Rev. Lett.},
  volume={71},
  number={9},
  pages={1291},
  year={1993},
  publisher={APS}, 
  doi={10.1103/PhysRevLett.71.1291}
}

@article{calmet2022quantumb,
  title={Quantum hair from gravity},
  author = {X. Calmet and R. Casadio and S. D. Hsu and F. Kuipers},
  journal={Phys. Rev. Lett.},
  volume={128},
  pages={111301},
  year={2022},
  publisher={APS}, 
  doi={10.1103/PhysRevLett.128.111301}
}

@article{papadodimas2014state,
  title={State-dependent bulk-boundary maps and black hole complementarity},
  author = {K. Papadodimas and S. Raju},
  journal={Phys. Rev.  D},
  volume={89},
  number={8},
  pages={086010},
  year={2014},
  publisher={APS},
  doi={10.1103/PhysRevD.89.086010}
}

@article{hawking2016soft,
  title={Soft hair on black holes},
  author = {S. W. Hawking and M. J. Perry and A. Strominger},
  journal={Phys. Rev. Lett.},
  volume={116},
  number={23},
  pages={231301},   DOI={10.1103/physrevlett.116.231301},
  year={2016},
  publisher={APS}
}

@inproceedings{preskill1992black,
  title={Do black holes destroy information},
  author = {J. Preskill},
  booktitle={Proc. Int. Symp. Black Holes, Membranes, Wormholes and Superstrings, S. Kalara and DV Nanopoulos, eds.(World Scientific, Singapore, 1993) pp},
  pages={22--39},
  year={1992},
  organization={World Scientific}
}

@article{liberati2019information,
  title={The information loss problem: an analogue gravity perspective},
  author = {S. Liberati and G. Tricella and A. Trombettoni},
  journal={Entropy},
  volume={21},
  number={10},
  pages={940},
  year={2019},
   DOI={10.3390/e21100940},
  publisher={MDPI}
}

@article{Eyheralde_2020,
   title={Quantum fluctuating geometries and the information paradox II},
   volume={37},
   ISSN={1361-6382},
   url={http://dx.doi.org/10.1088/1361-6382/ab6e89},
   DOI={10.1088/1361-6382/ab6e89},
   number={6},
   journal={Class. Quantum Gravity},
   publisher={IOP Publishing},
   author = {R. Eyheralde and R. Gambini and J. Pullin},
   year={2020},
   month=feb, pages={065001} }

@article{Eyheralde_2017,
   title={Quantum fluctuating geometries and the information paradox},
   volume={34},
   ISSN={1361-6382},
   url={http://dx.doi.org/10.1088/1361-6382/aa8e30},
   DOI={10.1088/1361-6382/aa8e30},
   number={23},
   journal={Class. Quantum Gravity},
   publisher={IOP Publishing},
   author = {R. Eyheralde and M. Campiglia and R. Gambini and J. Pullin},
   year={2017},
   month=nov, pages={235015} }

@article{brustein2013restoring,
  title={Restoring predictability in semiclassical gravitational collapse},
  author = {R. Brustein and A. Medved},
  journal={J. High Energy Phys.},
  volume={2013},
  number={9},
  pages={1--21},
  year={2013},   DOI={10.1007/jhep09(2013)015},
  publisher={Springer}
}

@article{alberte2015density,
  title={Density matrix of black hole radiation},
  author = {L. Alberte and R. Brustein and A. Khmelnitsky and A. Medved},
  journal={J. High Energy Phys.},
  volume={2015},
  number={8},
  pages={1--37},
  year={2015},
   DOI={10.1007/jhep08(2015)015},
  publisher={Springer}
}

@article{raju2022failure,
  title={Failure of the split property in gravity and the information paradox},
  author = {S. Raju},
  journal={Class. Quantum Gravity},
  volume={39},
  number={6},
  pages={064002},
  year={2022},
  publisher={IOP Publishing},
   DOI={10.1088/1361-6382/ac482b}
}

@article{almheiri2019entropy,
  title={The entropy of bulk quantum fields and the entanglement wedge of an evaporating black hole},
  author = {A. Almheiri and N. Engelhardt and D. Marolf and H. Maxfield},
  journal={J. High Energy Phys.},
  volume={2019},
  number={12},
  pages={1--47},
  year={2019},
  publisher={Springer},
  doi={10.1007/jhep12(2019)063}
}

@article{almheiri2020replica,
  title={Replica wormholes and the entropy of {H}awking radiation},
  author = {A. Almheiri and T. Hartman and J. Maldacena and E. Shaghoulian and A. Tajdini},
  journal={J. High Energy Phys.},
  volume={2020},
  number={5},
  pages={1--42},
  year={2020},
   DOI={10.1007/jhep05(2020)013},
  publisher={Springer}
}

@article{ng2022little,
  title={A little excitement across the horizon},
  author = {K. K. Ng and C. Zhang and J. Louko and R. B. Mann},
  journal={New J. Phys.},
  volume={24},
  number={10},
  pages={103018},
  year={2022},
   DOI={10.1088/1367-2630/ac9547},
  publisher={IOP Publishing}
}

@article{page2013time,
  title={Time dependence of {H}awking radiation entropy},
  author = {D. N. Page},
  journal={J. Cosmol. Astropart. Phys.},
  volume={2013},
  number={09},
  pages={028},
  year={2013},
  publisher={IOP Publishing},
   DOI={10.1088/1475-7516/2013/09/028}
}

@article{Uhlemann2022information,
   title={Information transfer with a twist},
   volume={2022},
   ISSN={1029-8479},
   url={http://dx.doi.org/10.1007/JHEP01(2022)126},
   DOI={10.1007/jhep01(2022)126},
   number={1},
   journal={J. High Energy Phys.},
   publisher={Springer Science and Business Media LLC},
   author={Uhlemann, C. F.},
   year={2022},
   month=jan }

@article{Uhlemann_2021,
   title={Islands and Page curves in 4d from Type IIB},
   volume={2021},
   ISSN={1029-8479},
   url={http://dx.doi.org/10.1007/JHEP08(2021)104},
   DOI={10.1007/jhep08(2021)104},
   number={8},
   journal={Journal of High Energy Physics},
   publisher={Springer Science and Business Media LLC},
   author={Uhlemann, Christoph F.},
   year={2021},
   month=aug }

@article{Chakraborty_2022,
   title={Monogamy paradox in empty flat space},
   volume={106},
   ISSN={2470-0029},
   url={http://dx.doi.org/10.1103/PhysRevD.106.086002},
   DOI={10.1103/physrevd.106.086002},
   pages={086002},
   journal={Phys. Rev. D},
   publisher={American Physical Society (APS)},
   author = {T. Chakraborty and J. Chakravarty and P. Paul},
   year={2022},
   month=oct }

@inproceedings{Polchinski_2016,
   title={The Black Hole Information Problem},
   url={http://dx.doi.org/10.1142/9789813149441_0006},
   DOI={10.1142/9789813149441_0006},
   booktitle={New Frontiers in Fields and Strings},
   publisher={World Scientific},
   author = {J. Polchinski},
   year={2016},
   month=nov }

@article{juarez-aubry_quantum_2023,
   title={Quantum strong cosmic censorship and black hole evaporation},
   volume={41},
   ISSN={1361-6382},
   url={http://dx.doi.org/10.1088/1361-6382/ad756c},
   DOI={10.1088/1361-6382/ad756c},
   number={19},
   journal={Class. Quantum Gravity},
   publisher={IOP Publishing},
   author = {B. A. Juárez-Aubry},
   year={2024},
   month=sep, pages={195027} }

@article{louko_hamiltonian_1998,
    title = {Hamiltonian spacetime dynamics with a spherical null-dust shell},
    volume = {57},
    copyright = {http://link.aps.org/licenses/aps-default-license},
    issn = {0556-2821, 1089-4918},
    url = {https://link.aps.org/doi/10.1103/PhysRevD.57.2279},
    doi = {10.1103/PhysRevD.57.2279},
    number = {4},
    urldate = {2025-04-21},
    journal = {Phys. Rev. D},
    author = {J. Louko and B. F. Whiting and J. L. Friedman},
    month = feb,
    year = {1998},
    pages = {2279--2298},
}

@article{gaddam_soft_2024,
    title = {Soft graviton exchange and the information paradox},
    volume = {109},
    issn = {2470-0010, 2470-0029},
    url = {https://link.aps.org/doi/10.1103/PhysRevD.109.026007},
    doi = {10.1103/PhysRevD.109.026007},
    number = {2},
    urldate = {2025-04-06},
    journal = {Phys. Rev. D},
    author = {N. Gaddam and N. Groenenboom},
    month = jan,
    year = {2024},
    pages = {026007},
}

@misc{walleghem2024stunnedsleepingbeautyprince,
      title={Stunned by Sleeping Beauty: How Prince Probability updates his forecast upon their fateful encounter}, 
      author = {L. Walleghem},
      year={2024},
      eprint={2408.06797},
      archivePrefix={arXiv},
      primaryClass={math.PR}}

@misc{hausmann2025firewallparadoxwignersfriend,
      title={The firewall paradox is {W}igner's friend paradox}, 
      author = {L. Hausmann and R. Renner},
      year={2025},
      eprint={2504.03835},
      archivePrefix={arXiv},
      primaryClass={quant-ph},
}

@misc{walleghem_refined_2024,
      title={A refined {F}rauchiger--{R}enner paradox based on strong contextuality}, 
      author = {L. Walleghem and R. S. Barbosa and M. Pusey and S. Weigert},
      year={2024},
      eprint={2409.05491},
      archivePrefix={arXiv},
      primaryClass={quant-ph}}

@misc{jones_thinking_2024,
      title={On the significance of Wigner's Friend in contexts beyond quantum foundations}, 
      author = {C. L. Jones and M. P. Mueller},
      year={2025},
      eprint={2402.08727},
      archivePrefix={arXiv},
      primaryClass={quant-ph}}

@article{cardoso2019testing,
  title={Testing the nature of dark compact objects: a status report},
  author = {V. Cardoso and P. Pani},
  journal={Living Rev. Relativ.},
  volume={22},
  pages={1--104},
  year={2019},
  publisher={Springer},
  doi={10.1007/s41114-019-0020-4}
}

@article{hayden2007black,
  title={Black holes as mirrors: quantum information in random subsystems},
  author = {P. Hayden and J. Preskill},
  journal={J. High Energy Phys.},
  volume={2007},
  number={09},
  pages={120},
  year={2007},
  publisher={IOP Publishing},
   DOI={10.1088/1126-6708/2007/09/120}
}

@article{steinhauer2016observation,
  title={Observation of quantum {H}awking radiation and its entanglement in an analogue black hole},
  author = {J. Steinhauer},
  journal={Nat. Phys.},
  volume={12},
  number={10},
  pages={959--965},
  year={2016},
   DOI={10.1038/nphys3863},
  publisher={Nature Publishing Group UK London}
}

@article{almheiri2020page,
  title={The {P}age curve of {H}awking radiation from semiclassical geometry},
  author = {A. Almheiri and R. Mahajan and J. Maldacena and Y. Zhao},
  journal={J. High Energy Phys.},
  volume={2020},
  number={3},
  pages={1--24},
  year={2020},
  publisher={Springer},
   DOI={10.1007/jhep03(2020)149}
}

@article{frauchiger2018quantum,
  title={Quantum theory cannot consistently describe the use of itself},
  author = {D. Frauchiger and R. Renner},
  journal={Nat. Commun.},
  volume={9},
  number={1},
  pages={3711},
  year={2018},
  publisher={Nature Publishing Group UK London},
  doi={10.1038/s41467-018-05739-8}
}

@article{wigner1995remarks,
  title={Remarks on the mind-body question},
  author = {E. P. Wigner},
  journal={Philosophical reflections and syntheses},
  pages={247--260},
  year={1995},
  publisher={Springer},
doi= {10.1007/978-3-642-78374-6_20}

}

@article{bell1964einstein,
  title={{On the {E}instein {P}odolsky {R}osen paradox}},
  author = {J. S. Bell},
  journal={Physics Physique Fizika},
  volume={1},
  number={3},
  pages={195},
  year={1964},
  publisher={APS}
}

@article{Bell_1976,
	title        = {{The Theory of Local Beables}},
	author = {J. S. Bell},
	year         = 1976,
	journal      = {Epistemological Letters},
	volume       = 9,
	pages        = {11--24},
	url          = {https://curate.nd.edu/show/c247dr29w4x}
}

@incollection{kochen1990onthe,
	doi = {10.1007/978-3-0348-9259-9_21},
	url = {https://doi.org/10.1007/978-3-0348-9259-9_21},
	year = 1990,
	publisher = {Birkhäuser Basel},
	pages = {235--263},
	author = {S. Kochen and E. P. Specker},
	title = {The Problem of Hidden Variables in Quantum Mechanics},
	booktitle = {Ernst Specker Selecta}
}

@misc{vilasini2022general,
  title={A general framework for consistent logical reasoning in {W}igner's friend scenarios: subjective perspectives of agents within a single quantum circuit},
  author = {V. Vilasini and M. P. Woods},
  archivePrefix={arXiv},
    eprint ={2209.09281},
    primaryClass = {quant-ph},
  year={2022}
}

@misc{susskind2013blackholecomplementarityharlowhayden,
      title={Black Hole Complementarity and the {H}arlow-{H}ayden Conjecture}, 
      author={L. Susskind},
      year={2013},
      eprint={1301.4505},
      archivePrefix={arXiv},
      primaryClass={hep-th}}

@article{BAIGUERA20261,
title = {Quantum complexity in gravity, quantum field theory, and quantum information science},
journal = {Physics Rep.},
volume = {1159},
pages = {1-77},
year = {2026},
issn = {0370-1573},
doi={10.1016/j.physrep.2025.11.001},
url = {https://www.sciencedirect.com/science/article/pii/S0370157325003023},
author = {S. Baiguera and V. Balasubramanian and P. Caputa and S. Chapman and J. Haferkamp and M. P. Heller and N. Y. Halpern}
}

@article{Verlinde_2013,
   title={Black hole entanglement and quantum error correction},
   volume={2013},
   ISSN={1029-8479},
   url={http://dx.doi.org/10.1007/JHEP10(2013)107},
   DOI={10.1007/jhep10(2013)107},
   number={10},
   journal={J. High Energy Phys.},
   publisher={Springer Science and Business Media LLC},
   author={Verlinde, E. and Verlinde, H.},
   year={2013},
   month=oct }

@article{Hayden_2019,
   title={Learning the Alpha-bits of black holes},
   volume={2019},
   ISSN={1029-8479},
   url={http://dx.doi.org/10.1007/JHEP12(2019)007},
   DOI={10.1007/jhep12(2019)007},
   number={12},
   journal={J. High Energy Phys.},
   publisher={Springer Science and Business Media LLC},
   author={Hayden, P. and Penington, G.},
   year={2019},
   month=dec }

@article{Pastawski_2015,
   title={Holographic quantum error-correcting codes: toy models for the bulk/boundary correspondence},
   volume={2015},
   ISSN={1029-8479},
   url={http://dx.doi.org/10.1007/JHEP06(2015)149},
   DOI={10.1007/jhep06(2015)149},
   pages={149},
   journal={J. High Energy Phys.},
   publisher={Springer Science and Business Media LLC},
   author={Pastawski, F. and Yoshida, B. and Harlow, D. and Preskill, J.},
   year={2015},
   month=jun }

@article{Almheiri_2015,
   title={Bulk locality and quantum error correction in {AdS/CFT}},
   volume={2015},
   ISSN={1029-8479},
   url={http://dx.doi.org/10.1007/JHEP04(2015)163},
   DOI={10.1007/jhep04(2015)163},
   pages={163},
   journal={J. High Energy Phys.},
   publisher={Springer Science and Business Media LLC},
   author={Almheiri, A. and Dong, X. and Harlow, D.},
   year={2015},
   month=apr }

@misc{almheiri2018holographicquantumerrorcorrection,
      title={Holographic Quantum Error Correction and the Projected Black Hole Interior}, 
      author={A. Almheiri},
      year={2018},
      eprint={1810.02055},
      archivePrefix={arXiv},
      primaryClass={hep-th}}

@article{Belin_2022,
   title={Does Complexity Equal Anything?},
   volume={128},
   ISSN={1079-7114},
   url={http://dx.doi.org/10.1103/PhysRevLett.128.081602},
   DOI={10.1103/physrevlett.128.081602},
   pages={081602},
   journal={Phys. Rev. Lett.},
   publisher={American Physical Society (APS)},
   author={Belin, A. and Myers, R. C. and Ruan, S-M. and Sárosi, G. and Speranza, A. J.},
   year={2022},
   month=feb }

@article{Brown_2017,
   title={Quantum complexity and negative curvature},
   volume={95},
   ISSN={2470-0029},
   url={http://dx.doi.org/10.1103/PhysRevD.95.045010},
   DOI={10.1103/physrevd.95.045010},
   pages={045010},
   journal={Phys. Rev. D},
   publisher={American Physical Society (APS)},
   author={Brown, A. R. and Susskind, L. and Zhao, Y.},
   year={2017},
   month=feb }

@article{Harlow_2014,
   title={Aspects of the {P}apadodimas-{R}aju proposal for the black hole interior},
   volume={2014},
   ISSN={1029-8479},
   url={http://dx.doi.org/10.1007/JHEP11(2014)055},
   DOI={10.1007/jhep11(2014)055},
   number={11},
   journal={J. High Energy Phys.},
   publisher={Springer Science and Business Media LLC},
   author={Harlow, Daniel},
   year={2014},
   month=nov }

@article{haddara2024local,
  title = {Local friendliness polytopes in multipartite scenarios},
  author = {M. Haddara and E. G. Cavalcanti},
  journal = {Phys. Rev. A},
  volume = {111},
  issue = {1},
  pages = {012206},
  numpages = {15},
  year = {2025},
  month = {Jan},
  publisher = {American Physical Society},
  doi = {10.1103/PhysRevA.111.012206},
  url = {https://link.aps.org/doi/10.1103/PhysRevA.111.012206}
}

@misc{walleghem2025extendedwignersfriendnogo,
      title={An extended Wigner's friend no-go theorem inspired by generalized contextuality}, 
      author = {L. Walleghem and L. Catani},
      year={2025},
      eprint={2502.02461},
      archivePrefix={arXiv},
      primaryClass={quant-ph}
}

@misc{strominger2018lecturesinfraredstructuregravity,
      title={Lectures on the Infrared Structure of Gravity and Gauge Theory}, 
      author = {A. Strominger},
      year={2018},
      eprint={1703.05448},
      archivePrefix={arXiv},
      primaryClass={hep-th}}

@article{An2023replica,
   title={Replica wormhole as a vacuum-to-vacuum transition},
   volume={83},
   ISSN={1434-6052},
   url={http://dx.doi.org/10.1140/epjc/s10052-023-11518-7},
   DOI={10.1140/epjc/s10052-023-11518-7},
   pages={341},
   journal={Eur. Phys. J. C},
   publisher={Springer Science and Business Media LLC},
   author = {Y. An and P. Cheng},
   year={2023},
   month=apr }

@article{eisert2010colloquium,
  title = {Colloquium: Area laws for the entanglement entropy},
  author = {J. Eisert and M. Cramer and M. B. Plenio},
  journal = {Rev. Mod. Phys.},
  volume = {82},
  issue = {1},
  pages = {277--306},
  numpages = {0},
  year = {2010},
  month = {Feb},
  publisher = {American Physical Society},
  doi = {10.1103/RevModPhys.82.277},
  url = {https://link.aps.org/doi/10.1103/RevModPhys.82.277}
}

@article{susskind1994gedanken,
  title = {Gedanken experiments involving black holes},
  author = {L. Susskind and L. Thorlacius},
  journal = {Phys. Rev. D},
  volume = {49},
  issue = {2},
  pages = {966--974},
  numpages = {0},
  year = {1994},
  month = {Jan},
  publisher = {American Physical Society},
  doi = {10.1103/PhysRevD.49.966},
  url = {https://link.aps.org/doi/10.1103/PhysRevD.49.966}
}

@article{vilasini2019multi,
  title={Multi-agent paradoxes beyond quantum theory},
  author = {V. Vilasini and N. Nurgalieva and L. del Rio},
  journal={New J. Phys.},
  volume={21},
  number={11},
  pages={113028},
  year={2019},
  publisher={IOP Publishing},
doi  = {10.1088/1367-2630/ab4fc4}
}

@article{brukner2018no,
  title={A no-go theorem for observer-independent facts},
  author = {{\v{C}}. Brukner},
  journal={Entropy},
  volume={20},
  number={5},
  pages={350},
  year={2018},
  publisher={MDPI},
 doi = {10.3390/e20050350}
}

@article{hardy1993nonlocality,
  title={Nonlocality for two particles without inequalities for almost all entangled states},
  author = {L. Hardy},
  journal={Phys. Rev. Lett.},
  volume={71},
  number={11},
  pages={1665},
  year={1993},
  publisher={APS},
doi = {10.1103/PhysRevLett.71.1665}
}

@misc{montanhano2023contextuality,
  title={Contextuality in multi-agent paradoxes},
  author = {S. B. Montanhano},
  archivePrefix={arXiv},
eprint = {2305.07792},
primaryClass = {quant-ph},
  year={2023}
}

@article{leegwater2022greenberger,
  title={When {G}reenberger, {H}orne and {Z}eilinger meet {W}igner's Friend},
  author = {G. Leegwater},
  journal={Foundations of Physics},
  volume={52},
  number={4},
  pages={68},
  year={2022},
  publisher={Springer},
doi = {10.1007/s10701-022-00586-6}
}

@article{bong2020strong,
  title={A strong no-go theorem on the {W}igner's friend paradox},
  author = {K. Bong and Utreras-Alarc{\'o}n, An{\'i}bal and Ghafari, Farzad and Liang, Yeong-Cherng and Tischler, Nora and Cavalcanti, Eric G and Pryde, Geoff J and Wiseman, Howard M},
  journal={Nat. Phys.},
  volume={16},
  number={12},
  pages={1199--1205},
  year={2020},
  publisher={Nature Publishing Group},
doi = {10.1038/s41567-020-0990-x}
}

@misc{cavalcanti2023consistency,
  title={On the consistency of relative facts},
  author = {E. G. Cavalcanti and A. Di Biagio and C. Rovelli},
  primaryClass={quant-ph}, archivePrefix={arXiv}, eprint={2305.07343},
  year={2023}
}

@article{wiseman2022thoughtful,
   title={A “thoughtful” Local Friendliness no-go theorem: a prospective experiment with new assumptions to suit},
   volume={7},
   ISSN={2521-327X},
   url={http://dx.doi.org/10.22331/q-2023-09-14-1112},
   DOI={10.22331/q-2023-09-14-1112},
   journal={Quantum},
   publisher={Verein zur Forderung des Open Access Publizierens in den Quantenwissenschaften},
   author = {H. M. Wiseman and E. G. Cavalcanti and E. G. Rieffel},
   year={2023},
   month=sep, pages={1112} }

@article{cavalcanti2021implications,
  title={Implications of {L}ocal {F}riendliness violation for quantum causality},
  author = {E. G. Cavalcanti and H. M. Wiseman},
  journal={Entropy},
  volume={23},
  number={8},
  pages={925},
  year={2021},
  publisher={Multidisciplinary Digital Publishing Institute},
doi = {10.3390/e23080925}
}

@misc{carroll2017boltzmannbrainsbad,
      title={Why {B}oltzmann Brains Are Bad}, 
      author = {S. M. Carroll},
      year={2017},
      eprint={1702.00850},
      archivePrefix={arXiv},
      primaryClass={hep-th}
}

@article{Steinhauer_2014,
   title={Observation of self-amplifying {H}awking radiation in an analogue black-hole laser},
   volume={10},
   ISSN={1745-2481},
   url={http://dx.doi.org/10.1038/NPHYS3104},
   DOI={10.1038/nphys3104},
   number={11},
   journal={Nat. Phys.},
   publisher={Springer Science and Business Media LLC},
   author = {J. Steinhauer},
   year={2014},
   month=oct, pages={864–869} }

@article{lahav2014realisation,
  title = {Realization of a Sonic Black Hole Analog in a {B}ose--{E}instein Condensate},
  author = {O. Lahav and A. Itah and A. Blumkin and C. Gordon and S. Rinott and A. Zayats and J. Steinhauer},
  journal = {Phys. Rev. Lett.},
  volume = {105},
  issue = {24},
  pages = {240401},
  numpages = {4},
  year = {2010},
  month = {Dec},
  publisher = {American Physical Society},
  doi = {10.1103/PhysRevLett.105.240401},
  url = {https://link.aps.org/doi/10.1103/PhysRevLett.105.240401}
}

@article{bose2025massive,
  title = {Massive quantum systems as interfaces of quantum mechanics and gravity},
  author = {S. Bose and I. Fuentes and A. A. Geraci and S. M. Khan and S. Qvarfort and M. Rademacher and M. Rashid and Toro\ifmmode \check{s}\else \v{s}\fi{}, M. and Ulbricht, H. and Wanjura, C. C.},
  journal = {Rev. Mod. Phys.},
  volume = {97},
  issue = {1},
  pages = {015003},
  numpages = {71},
  year = {2025},
  month = {Feb},
  publisher = {American Physical Society},
  doi = {10.1103/RevModPhys.97.015003},
  url = {https://link.aps.org/doi/10.1103/RevModPhys.97.015003}
}

@article{Munoz2019observation,
   title={Observation of thermal {H}awking radiation and its temperature in an analogue black hole},
   volume={569},
   ISSN={1476-4687},
   url={http://dx.doi.org/10.1038/s41586-019-1241-0},
   DOI={10.1038/s41586-019-1241-0},
   number={7758},
   journal={Nature},
   publisher={Springer Science and Business Media LLC},
   author = {J. R. Muñoz de Nova and K. Golubkov and V. I. Kolobov and J. Steinhauer},
   year={2019},
   month=may, pages={688–691} }

@article{drori2019observation,
  title = {Observation of Stimulated {H}awking Radiation in an Optical Analogue},
  author = {J. Drori and Y. Rosenberg and D. Bermudez and Y. Silberberg and U. Leonhardt},
  journal = {Phys. Rev. Lett.},
  volume = {122},
  issue = {1},
  pages = {010404},
  numpages = {6},
  year = {2019},
  month = {Jan},
  publisher = {American Physical Society},
  doi = {10.1103/PhysRevLett.122.010404},
  url = {https://link.aps.org/doi/10.1103/PhysRevLett.122.010404}
}

@article{unruh1995sonic,
  title = {Sonic analogue of black holes and the effects of high frequencies on black hole evaporation},
  author = {W. G. Unruh},
  journal = {Phys. Rev. D},
  volume = {51},
  issue = {6},
  pages = {2827--2838},
  numpages = {0},
  year = {1995},
  month = {Mar},
  publisher = {American Physical Society},
  doi = {10.1103/PhysRevD.51.2827},
  url = {https://link.aps.org/doi/10.1103/PhysRevD.51.2827}
}

@misc{schmid2023review,
      title={A review and analysis of six extended Wigner's friend arguments}, 
      author = {D. Schmid and Y. Yīng and M. Leifer},
      year={2024},
      eprint={2308.16220},
      archivePrefix={arXiv},
      primaryClass={quant-ph}
}

@article{haddara2022possibilistic,
   title={A possibilistic no-go theorem on the {W}igner’s friend paradox},
  author = {M. Haddara and E. G. Cavalcanti},
  journal={New J. Phys.},
  volume={25},
  number={9},
  pages={093028},
  year={2023},
doi = {10.1088/1367-2630/aceea3}
}

@misc{ying2023relating,
      title={Relating Wigner's Friend scenarios to Nonclassical Causal Compatibility, Monogamy Relations, and Fine Tuning}, 
      author = {Y. Yīng and M. M. Ansanelli and A. D. Biagio and E. Wolfe and E. G. Cavalcanti},
      year={2023},
      eprint={2309.12987},
      archivePrefix={arXiv},
      primaryClass={quant-ph}}

@misc{ormrod2022no,
  title={A no-go theorem for absolute observed events without inequalities or modal logic},
  author = {N. Ormrod and J. Barrett},
  archivePrefix={arXiv},
eprint = {2209.03940},
primaryClass = {quant-ph},
  year={2022}
}

@misc{ormrod2023theories,
  title={Which theories have a measurement problem?},
  author = {N. Ormrod and V. Vilasini and J. Barrett},
  archivePrefix={arXiv},
eprint={2303.03353},
primaryClass = {quant-ph},
  year={2023}
}

@misc{nurgalieva2018inadequacy,
  title={Inadequacy of modal logic in quantum settings},
  author = {N. Nurgalieva and L. del Rio},
  archivePrefix={arXiv},
eprint = {1804.01106},
primaryClass = {quant-ph},
  year={2018}
}

@article{zukowski2021physics,
  title={Physics and metaphysics of {W}igner's friends: {E}ven performed premeasurements have no results},
  author = {{\.Z}ukowski, Marek and Markiewicz, Marcin},
  journal={Phys. Rev. Lett.},
  volume={126},
  number={13},
  pages={130402},
  year={2021},
  publisher={APS},
doi = {10.1103/PhysRevLett.126.130402}
}

@article{walleghem2023extended,
  title={Extended {W}igner's friend paradoxes do not require nonlocal correlations},
  author = {L. Walleghem and R. Wagner and D. Schmid and Y{\=i}ng, Y{\`i}l{\`e}},
  volume={112},
   ISSN={2469-9934},
   url={http://dx.doi.org/10.1103/n4hv-rlgj},
   DOI={10.1103/n4hv-rlgj},
   pages={022212},
   journal={Phys. Rev. A},
   publisher={American Physical Society (APS)},
   year={2025}
}

@article{guerin2020no,
  title={A no-go theorem for the persistent reality of {W}igner's friend's perception},
  author = {Gu{\'e}rin, P. A. and Baumann, V. and Del Santo, F. and Brukner, {\v{C}}.},
  journal={Commun. Phys.},
  volume={4},
  number={1},
  pages={93},
  year={2021},
  publisher={Nature Publishing Group UK London},
doi = {0.1038/s42005-021-00589-1}
}

@article{brukner2017quantum,
  title={On the quantum measurement problem},
  author = {{\v{C}}. Brukner},
  journal={Quantum [Un] Speakables II: Half a Century of Bell's Theorem},
  pages={95--117},
  year={2017},
  publisher={Springer},
doi = {10.48550/arXiv.1507.05255}
}

@misc{szangolies2020quantum,
  title={The Quantum {R}ashomon Effect: {A} Strengthened {F}rauchiger-{R}enner Argument},
  author = {J. Szangolies},
  primaryClass={quant-ph}, archivePrefix={arXiv}, eprint={2011.12716},
  year={2020}
}

@misc{utreras2023allowing,
  title={Allowing {W}igner's friend to sequentially measure incompatible observables},
  author = {Utreras-Alarc{\'o}n, An{\'i}bal and Cavalcanti, Eric G and Wiseman, Howard M},
  primaryClass={quant-ph}, archivePrefix={arXiv}, eprint={2305.09102},
  year={2023}
}

@article{abramsky2011sheaf,
  title={The sheaf-theoretic structure of non-locality and contextuality},
  author = {S. Abramsky and A. Brandenburger},
  journal={New J. Phys.},
  volume={13},
  number={11},
  pages={113036},
  year={2011},
  publisher={IOP Publishing},
 doi ={10.1088/1367-2630/13/11/113036}
}

@phdthesis{walleghem2026thesis,
  title={Foundational puzzles on quantum
universality: From {W}igner’s friend to black holes},
  author = {L. Walleghem},
  year={2026},
  type = {{DP}hil thesis},
  school={University of York}
}

@article{oreshkov2012quantum,
  title={Quantum correlations with no causal order},
  author = {O. Oreshkov and F. Costa and {\v{C}}. Brukner},
  journal={Nat. Commun.},
  volume={3},
  number={1},
  pages={1092},
  year={2012},
  publisher={Nature Publishing Group UK London},
 doi = {10.1038/ncomms2076}
}

@article{debrota2020respecting,
  title={Respecting one’s fellow: {QB}ism’s analysis of {W}igner’s friend},
  author = {J. B. DeBrota and C. A. Fuchs and R. Schack},
  journal={Foundations of Physics},
  volume={50},
  pages={1859--1874},
  year={2020},
  publisher={Springer},
  doi={10.1007/s10701-020-00369-x}
}

@misc{brukner2021qubits,
  title={Qubits are not observers--a no-go theorem},
  author = {{\v{C}}. Brukner},
  primaryClass={quant-ph}, archivePrefix={arXiv}, eprint={2107.03513},
  year={2021}
}

@article{schlosshauer2005decoherence,
  title = {Decoherence, the measurement problem, and interpretations of quantum mechanics},
  author = {M. Schlosshauer},
  journal = {Rev. Mod. Phys.},
  volume = {76},
  issue = {4},
  pages = {1267--1305},
  numpages = {0},
  year = {2005},
  publisher = {American Physical Society},
  doi = {10.1103/RevModPhys.76.1267},
  url = {https://link.aps.org/doi/10.1103/RevModPhys.76.1267}
}

@article{rovelli1996relational,
  title={Relational quantum mechanics},
  author = {C. Rovelli},
  journal={Int. J. Theor. Phys.},
  volume={35},
  pages={1637--1678},
  year={1996},
  publisher={Springer},
  doi={10.1007/BF02302261}
}

@article{rovelli2018space,
  title={Space is blue and birds fly through it},
  author = {C. Rovelli},
  journal={Phil. Trans. R. Soc. A.},
  volume={376},
  number={2123},
  pages={20170312},
  year={2018},
  publisher={The Royal Society Publishing},
  doi ={10.1098/rsta.2017.0312}
}

@article{walleghem2024connecting,
  title = {Connecting extended {W}igner’s friend arguments and noncontextuality},
  author = {L. Walleghem and Y{\=i}ng, Y{\`i}l{\`e} and Wagner, Rafael and Schmid, David},
  volume={9},
   ISSN={2521-327X},
   url={http://dx.doi.org/10.22331/q-2025-07-31-1819},
   DOI={10.22331/q-2025-07-31-1819},
   journal={Quantum},
   publisher={Verein zur Forderung des Open Access Publizierens in den Quantenwissenschaften},
   year={2025},
   month=jul, pages={1819} }

@article{spekkens2005contextuality,
  title={Contextuality for preparations, transformations, and unsharp measurements},
  author = {R. W. Spekkens},
  journal={Phys. Rev. A},
  volume={71},
  number={5},
  pages={052108},
  year={2005},
  publisher={APS},
  doi = {10.1103/PhysRevA.71.052108}
}

@article{schmid2020structure,
  title={A structure theorem for generalized-noncontextual ontological models},
  author = {D. Schmid and J. H. Selby and M. F. Pusey and R. W. Spekkens},
  journal = {Quantum},
    volume ={8}, 
pages = {1283},
  year={2024},
doi ={10.22331/q-2024-03-14-1283}
}

@article{brunner_bell_2014,
    title = {Bell nonlocality},
    volume = {86},
    issn = {0034-6861, 1539-0756},
    url = {http://arxiv.org/abs/1303.2849},
    doi = {10.1103/RevModPhys.86.419},
    number = {2},
    urldate = {2023-02-12},
    journal = {Rev. Modern Phys.},
    author = {N. Brunner and D. Cavalcanti and S. Pironio and V. Scarani and S. Wehner},
    month = apr,
    year = {2014},
    note = {arXiv:1303.2849 [quant-ph]},
    pages = {419--478},
}

@article{budroni_kochen-specker_2022,
    title = {Kochen-{Specker} contextuality},
    volume = {94},
    issn = {0034-6861, 1539-0756},
    url = {https://link.aps.org/doi/10.1103/RevModPhys.94.045007},
    doi = {10.1103/RevModPhys.94.045007},
    number = {4},
    urldate = {2023-06-06},
    journal = {Rev. Modern Phys.},
    author = {C. Budroni and A. Cabello and O. Gühne and M. Kleinmann and J. Larsson},
    month = dec,
    year = {2022},
    pages = {045007},
}

@article{foo2021entanglement,
  title = {Entanglement amplification between superposed detectors in flat and curved spacetimes},
  author = {J. Foo and R. B. Mann and M. Zych},
  journal = {Phys. Rev. D},
  volume = {103},
  issue = {6},
  pages = {065013},
  numpages = {19},
  year = {2021},
  month = {Mar},
  publisher = {American Physical Society},
  doi = {10.1103/PhysRevD.103.065013},
  url = {https://link.aps.org/doi/10.1103/PhysRevD.103.065013}
}

@article{fine1982joint,
    author = {A. Fine},
    title = {Joint distributions, quantum correlations, and commuting observables},
    journal = {J. Math. Phys.},
    volume = {23},
    number = {7},
    pages = {1306-1310},
    year = {1982},
    month = {07},
    issn = {0022-2488},
    doi = {10.1063/1.525514},
    url = {https://doi.org/10.1063/1.525514}
}

@misc{calmet2024blackholeinformationreplica,
      title={Black Hole Information, Replica Wormholes, and Macroscopic Entanglement}, 
      author = {X. Calmet and S. D. H. Hsu},
      year={2024},
      eprint={2412.07807},
      archivePrefix={arXiv},
      primaryClass={hep-th},
}

@article{Alsing_2012,
   title={Observer-dependent entanglement},
   volume={29},
   ISSN={1361-6382},
   url={http://dx.doi.org/10.1088/0264-9381/29/22/224001},
   DOI={10.1088/0264-9381/29/22/224001},
   number={22},
   journal={Class. Quantum Gravity},
   publisher={IOP Publishing},
   author = {P. M. Alsing and I. Fuentes},
   year={2012},
   month=oct, pages={224001} }

@article{fine1982hidden,
  title = {Hidden Variables, Joint Probability, and the Bell Inequalities},
  author = {A. Fine},
  journal = {Phys. Rev. Lett.},
  volume = {48},
  issue = {5},
  pages = {291--295},
  numpages = {0},
  year = {1982},
  month = {Feb},
  publisher = {American Physical Society},
  doi = {10.1103/PhysRevLett.48.291},
  url = {https://link.aps.org/doi/10.1103/PhysRevLett.48.291}
}

@article{Oppenheim_2014,
   title={Firewalls and flat mirrors: An alternative to the AMPS experiment which evades the Harlow-Hayden obstacle},
   volume={2014},
   ISSN={1029-8479},
   url={http://dx.doi.org/10.1007/JHEP03(2014)120},
   DOI={10.1007/jhep03(2014)120},
   number={3},
   journal={J. High Energy Phys.},
   publisher={Springer Science and Business Media LLC},
   author = {J. Oppenheim and B. Unruh},
   year={2014},
   month=mar }

@article{PhysRevResearch.3.043056,
  title = {Thermality, causality, and the quantum-controlled {U}nruh--{DeWitt} detector},
  author = {J. Foo and S. Onoe and R. B. Mann and M. Zych},
  journal = {Phys. Rev. Res.},
  volume = {3},
  issue = {4},
  pages = {043056},
  numpages = {9},
  year = {2021},
  month = {Oct},
  publisher = {American Physical Society},
  doi = {10.1103/PhysRevResearch.3.043056},
  url = {https://link.aps.org/doi/10.1103/PhysRevResearch.3.043056}
}

@misc{chakraborty2024entanglementharvestingquantumsuperposed,
      title={Entanglement harvesting in quantum superposed spacetime}, 
      author = {A. Chakraborty and L. Hackl and M. Zych},
      year={2024},
      eprint={2412.15870},
      archivePrefix={arXiv},
      primaryClass={gr-qc},
}

@article{carney_dressed_2018,
    title = {Dressed infrared quantum information},
    volume = {97},
    issn = {2470-0010, 2470-0029},
    url = {https://link.aps.org/doi/10.1103/PhysRevD.97.025007},
    doi = {10.1103/PhysRevD.97.025007},
    
    number = {2},
    urldate = {2025-05-08},
    journal = {Phys. Rev. D},
    author = {D. Carney and L. Chaurette and D. Neuenfeld and G. W. Semenoff},
    month = jan,
    year = {2018},
    pages = {025007},
}

@article{hoehn_trinity_2021,
    title = {The {Trinity} of {Relational} {Quantum} {Dynamics}},
    volume = {104},
    issn = {2470-0010, 2470-0029},
    doi = {10.1103/PhysRevD.104.066001},
    number = {6},
    urldate = {2025-05-07},
    journal = {Phys. Rev. D},
    author = {P. A. Hoehn and A. R. H. Smith and M. P. E. Lock},
    month = sep,
    year = {2021},
    pages = {066001},
}

@article{giacomini_spacetime_2021,
    title = {Spacetime {Quantum} {Reference} {Frames} and superpositions of proper times},
    volume = {5},
    issn = {2521-327X},
    doi = {10.22331/q-2021-07-22-508},    
    urldate = {2025-04-22},
    journal = {Quantum},
    author = {F. Giacomini},
    month = jul,
    year = {2021},
    pages = {508},
}

@article{vanrietvelde_switching_2023,
    title = {Switching quantum reference frames in the {N}-body problem and the absence of global relational perspectives},
    volume = {7},
    issn = {2521-327X},
    doi = {10.22331/q-2023-08-22-1088},
    urldate = {2025-05-07},
    journal = {Quantum},
    author = {A. Vanrietvelde and P. A. Hoehn and F. Giacomini},
    month = aug,
    year = {2023},
    pages = {1088},
}

@article{rovelli_quantum_1990,
    title = {Quantum mechanics without time: {A} model},
    volume = {42},
    copyright = {http://link.aps.org/licenses/aps-default-license},
    issn = {0556-2821},
    shorttitle = {Quantum mechanics without time},
    url = {https://link.aps.org/doi/10.1103/PhysRevD.42.2638},
    doi = {10.1103/PhysRevD.42.2638},
    
    number = {8},
    urldate = {2025-05-06},
    journal = {Phys. Rev. D},
    author = {C. Rovelli},
    month = oct,
    year = {1990},
    pages = {2638--2646},
}

@article{holzhey_geometric_1994,
    title = {Geometric and renormalized entropy in conformal field theory},
    volume = {424},
    copyright = {https://www.elsevier.com/tdm/userlicense/1.0/},
    issn = {05503213},
    url = {https://linkinghub.elsevier.com/retrieve/pii/0550321394904022},
    doi = {10.1016/0550-3213(94)90402-2},    
    number = {3},
    urldate = {2025-06-01},
    journal = {Nucl. Phys. B},
    author = {C. Holzhey and F. Larsen and F. Wilczek},
    month = aug,
    year = {1994},
    pages = {443--467},
}

@article{flanagan_order-unity_2021,
    title = {Order-{Unity} {Correction} to {Hawking} {Radiation}},
    volume = {127},
    issn = {0031-9007, 1079-7114},
    url = {https://link.aps.org/doi/10.1103/PhysRevLett.127.041301},
    doi = {10.1103/PhysRevLett.127.041301},
    
    number = {4},
    urldate = {2025-04-06},
    journal = {Phys. Rev. Lett.},
    author = {E. E. Flanagan},
    month = jul,
    year = {2021},
    pages = {041301},
}

@article{ashtekar_surprises_2011,
  title = {Surprises in the Evaporation of 2D Black Holes},
  author = {A. Ashtekar and F. Pretorius and Ramazanoğlu, F. M.},
  journal = {Phys. Rev. Lett.},
  volume = {106},
  issue = {16},
  pages = {161303},
  numpages = {4},
  year = {2011},
  month = {Apr},
  publisher = {American Physical Society},
  doi = {10.1103/PhysRevLett.106.161303},
  url = {https://link.aps.org/doi/10.1103/PhysRevLett.106.161303}
}

@article{harlow_quantum_2013,
    title = {Quantum {Computation} vs. {Firewalls}},
    volume = {2013},
    issn = {1029-8479},
    url = {http://arxiv.org/abs/1301.4504},
    doi = {10.1007/JHEP06(2013)085},
    number = {6},
    urldate = {2025-06-03},
    journal = {J. High Energy Phys.},
    author = {D. Harlow and P. Hayden},
    month = jun,
    year = {2013},
    note = {arXiv:1301.4504 [hep-th]},
    pages = {85},
}

@misc{aaronson_complexity_2016,
      title={The Complexity of Quantum States and Transformations: From Quantum Money to Black Holes}, 
      author = {S. Aaronson},
      year={2016},
      eprint={1607.05256},
      archivePrefix={arXiv},
      primaryClass={quant-ph},
}

@article{pasterski_hps_2021,
    title = {{HPS} meets {AMPS}: {How} {Soft} {Hair} {Dissolves} the {Firewall}},
    volume = {2021},
    issn = {1029-8479},
    shorttitle = {{HPS} meets {AMPS}},
    doi = {10.1007/JHEP09(2021)099},
    number = {9},
    urldate = {2025-04-25},
    journal = {J. High Energy Phys.},
    author = {S. Pasterski and H. Verlinde},
    month = sep,
    year = {2021},
    pages = {99},
}

@article{kim_ghost_2020,
    title = {The ghost in the radiation: {Robust} encodings of the black hole interior},
    volume = {2020},
    issn = {1029-8479},
    shorttitle = {The ghost in the radiation},
    url = {http://arxiv.org/abs/2003.05451},
    doi = {10.1007/JHEP06(2020)031},
    number = {6},
    urldate = {2025-06-11},
    journal = {J. High Energy Phys.},
    author = {I. H. Kim and E. Tang and J. Preskill},
    month = jun,
    year = {2020},
    note = {arXiv:2003.05451 [hep-th]},
    pages = {31},
}

@article{ashtekar_information_2008,
    title = {Information is {Not} {Lost} in the {Evaporation} of 2-dimensional {Black} {Holes}},
    volume = {100},
    issn = {0031-9007, 1079-7114},
    url = {http://arxiv.org/abs/0801.1811},
    doi = {10.1103/PhysRevLett.100.211302},
    number = {21},
    urldate = {2025-04-06},
    journal = {Phys. Rev. Lett.},
    author = {A. Ashtekar and V. Taveras and M. Varadarajan},
    month = may,
    year = {2008},
    pages = {211302},
}

@article{yoshida_firewalls_2019,
    title = {Firewalls vs. {Scrambling}},
    volume = {2019},
    issn = {1029-8479},
    url = {http://arxiv.org/abs/1902.09763},
    doi = {10.1007/JHEP10(2019)132},
    number = {10},
    urldate = {2025-06-11},
    journal = {J. High Energy Phys.},
    author = {B. Yoshida},
    month = oct,
    year = {2019},
    note = {arXiv:1902.09763 [hep-th]},
    pages = {132},
}

@misc{antonini2025apologiaislands,
      title={An apologia for islands}, 
      author = {S. Antonini and C. Chen and H. Maxfield and G. Penington},
      year={2025},
      eprint={2506.04311},
      archivePrefix={arXiv},
      primaryClass={hep-th},
}

@article{Krishnan_2021,
   title={Critical islands},
   volume={2021},
   ISSN={1029-8479},
   url={http://dx.doi.org/10.1007/JHEP01(2021)179},
   DOI={10.1007/jhep01(2021)179},
   number={1},
   journal={J. High Energy Phys.},
   publisher={Springer Science and Business Media LLC},
   author = {C. Krishnan},
   year={2021},
   month=jan }

@misc{bousso_islands_2023,
    title = {Islands far outside the horizon},
year={2023},
      eprint={2312.03078},
      archivePrefix={arXiv},
      primaryClass={hep-th},
    author = {R. Bousso and G. Penington},
    }

@article{carney_infrared_2017,
    title = {Infrared quantum information},
    volume = {119},
    issn = {0031-9007, 1079-7114},
    url = {http://arxiv.org/abs/1706.03782},
    doi = {10.1103/PhysRevLett.119.180502},   
    number = {18},
    journal = {Phys. Rev. Lett.},
    author = {D. Carney and L. Chaurette and D. Neuenfeld and G. W. Semenoff},
    month = oct,
    year = {2017},
    note = {arXiv:1706.03782 [hep-th]},
    pages = {180502},
}

@article{bondi1962gravitational,
  title={Gravitational waves in general relativity, VII. Waves from axi-symmetric isolated system},
  author = {H. Bondi and M. G. J. Van der Burg and A. Metzner},
  journal={Proc. R. Soc. London, Ser. A},
  volume={269},
  number={1336},
  pages={21--52},
  year={1962},
  publisher={The Royal Society London},
doi = {10.1098/rspa.1962.0161}
}

@article{sachs1962gravitational,
  title={Gravitational waves in general relativity VIII. Waves in asymptotically flat space-time},
  author = {R. K. Sachs},
  journal={Proc. R. Soc. London, Ser. A},
  volume={270},
  number={1340},
  pages={103--126},
  year={1962},
  publisher={The Royal Society London}, 
  doi = {10.1098/rspa.1962.0206}
}

@article{flanagan_conserved_2017,
    title = {Conserved charges of the extended {Bondi}-{Metzner}-{Sachs} algebra},
    volume = {95},
    copyright = {http://link.aps.org/licenses/aps-default-license},
    issn = {2470-0010, 2470-0029},
    url = {https://link.aps.org/doi/10.1103/PhysRevD.95.044002},
    doi = {10.1103/PhysRevD.95.044002},
    
    number = {4},
    journal = {Phys. Rev. D},
    author = {E. E. Flanagan and D. A. Nichols},
    month = feb,
    year = {2017},
    pages = {044002},
}

@article{compere_asymptotic_2023,
    title = {An asymptotic framework for gravitational scattering},
    volume = {40},
    issn = {0264-9381, 1361-6382},
    url = {https://iopscience.iop.org/article/10.1088/1361-6382/acf5c1},
    doi = {10.1088/1361-6382/acf5c1},
    
    number = {20},
    urldate = {2025-08-27},
    journal = {Class. Quantum Gravity},
    author = {G. Compère and S. E. Gralla and H. Wei},
    month = oct,
    year = {2023},
    pages = {205018},
}

@article{compere_poincare_2020,
    title = {The {Poincaré} and {BMS} flux-balance laws with application to binary systems},
    volume = {2020},
    issn = {1029-8479},
    doi = {10.1007/JHEP10(2020)116},
    
    number = {10},
    journal = {J. High Energy Phys.},
    author = {G. Compère and R. Oliveri and A. Seraj},
    month = oct,
    year = {2020},
    pages = {116},
}

@misc{compère2019advancedlecturesgeneralrelativity,
      title={Advanced Lectures on General Relativity}, 
      author = {G. Compère and A. Fiorucci},
      year={2019},
      eprint={1801.07064},
      archivePrefix={arXiv},
      primaryClass={hep-th}}

@article{satishchandran2019asymptotic,
  title = {Asymptotic behavior of massless fields and the memory effect},
  author = {G. Satishchandran and R. M. Wald},
  journal = {Phys. Rev. D},
  volume = {99},
  issue = {8},
  pages = {084007},
  numpages = {38},
  year = {2019},
  month = {Apr},
  publisher = {American Physical Society},
  doi = {10.1103/PhysRevD.99.084007},
  url = {https://link.aps.org/doi/10.1103/PhysRevD.99.084007}
}

@article{prabhu_infrared_2024,
    title = {Infrared finite scattering theory: {Amplitudes} and soft theorems},
    volume = {110},
    issn = {2470-0010, 2470-0029},
    shorttitle = {Infrared finite scattering theory},
    url = {https://link.aps.org/doi/10.1103/PhysRevD.110.085022},
    doi = {10.1103/PhysRevD.110.085022},
    
    number = {8},
    journal = {Phys. Rev. D},
    author = {K. Prabhu and G. Satishchandran},
    month = oct,
    year = {2024},
    pages = {085022},
}

@article{Chowdhury_2022,
   title={Holography from the {Wheeler}--{DeWitt} equation},
   volume={2022},
   ISSN={1029-8479},
   url={http://dx.doi.org/10.1007/JHEP03(2022)019},
   DOI={10.1007/jhep03(2022)019},
   number={3},
   journal={J. High Energy Phys.},
   publisher={Springer Science and Business Media LLC},
   author = {C. Chowdhury and V. Godet and O. Papadoulaki and S. Raju},
   year={2022},
   month=mar }

@article{Bao_2018,
   title={Branches of the black hole wave function need not contain firewalls},
   volume={97},
   ISSN={2470-0029},
   url={http://dx.doi.org/10.1103/PhysRevD.97.126014},
   DOI={10.1103/physrevd.97.126014},
   pages={126014},
   journal={Phys. Rev. D},
   publisher={American Physical Society (APS)},
   author = {N. Bao and S. M. Carroll and A. Chatwin-Davies and J. Pollack and G. N. Remmen},
   year={2018},
   month=jun }

@misc{akil2025quantumsuperpositionblackhole,
      title={A Quantum Superposition of Black Hole Evaporation Histories: Recovering Unitarity}, 
      author = {A. Akil and L. Giannelli and L. Modesto and O. Dahlsten and G. Chiribella and {\v{C}}. Brukner},
      year={2025},
      eprint={2507.17031},
      archivePrefix={arXiv},
      primaryClass={gr-qc}}

@article{flanagan_infrared_2021,
    title = {Infrared {Effects} in the {Late} {Stages} of {Black} {Hole} {Evaporation}},
    volume = {2021},
    issn = {1029-8479},
    url = {http://arxiv.org/abs/2102.13629},
    doi = {10.1007/JHEP07(2021)137},
    
    number = {7},
    urldate = {2025-06-18},
    journal = {J. High Energy Phys.},
    author = {E. E. Flanagan},
    month = jul,
    year = {2021},
    note = {arXiv:2102.13629 [hep-th]},
    pages = {137},
}

@article{Donnay_2022carrollian,
   title={Carrollian Perspective on Celestial Holography},
   volume={129},
   ISSN={1079-7114},
   url={http://dx.doi.org/10.1103/PhysRevLett.129.071602},
   DOI={10.1103/physrevlett.129.071602},
   number={7},
    pages={071602},
   journal={Phys. Rev. Lett.},
   publisher={American Physical Society (APS)},
   author = {L. Donnay and A. Fiorucci and Y. Herfray and R. Ruzziconi},
   year={2022} }

@article{Araujo_Regado_2023,
   title={Cauchy slice holography: a new {AdS/CFT} dictionary},
   volume={2023},
   ISSN={1029-8479},
   url={http://dx.doi.org/10.1007/JHEP03(2023)026},
   DOI={10.1007/jhep03(2023)026},
   number={3},
   journal={J. High Energy Phys.},
   publisher={Springer Science and Business Media LLC},
   author = {G. Araujo-Regado and R. Khan and A. C. Wall},
   year={2023},
   month=mar }

@misc{donnay_celestial_2023,
      title={Celestial holography: An asymptotic symmetry perspective}, 
      author = {L. Donnay},
      year={2023},
      eprint={2310.12922},
      archivePrefix={arXiv},
      primaryClass={hep-th}}

@misc{hooft2009dimensionalreductionquantumgravity,
      title={Dimensional Reduction in Quantum Gravity}, 
      author = {G. 't Hooft},
      year={1993},
      eprint={gr-qc/9310026},
      archivePrefix={arXiv},
      primaryClass={gr-qc}}

@article{Susskind_1995,
   title={The world as a hologram},
   volume={36},
   ISSN={1089-7658},
   url={http://dx.doi.org/10.1063/1.531249},
   DOI={10.1063/1.531249},
   number={11},
   journal={J. Math. Phys.},
   publisher={AIP Publishing},
   author = {L. Susskind},
   year={1995},
   month=nov, pages={6377–6396} }

@article{Schindler_2020,
   title={Understanding black hole evaporation using explicitly computed Penrose diagrams},
   volume={101},
   ISSN={2470-0029},
   url={http://dx.doi.org/10.1103/PhysRevD.101.024010},
   DOI={10.1103/physrevd.101.024010},
   pages={024010},
   journal={Phys. Rev. D},
   publisher={American Physical Society (APS)},
   author = {J. C. Schindler and A. Aguirre and A. Kuttner},
   year={2020},
   month=jan }

@article{Ashtekar_2005,
   title={Black hole evaporation: a paradigm},
   volume={22},
   ISSN={1361-6382},
   url={http://dx.doi.org/10.1088/0264-9381/22/16/014},
   DOI={10.1088/0264-9381/22/16/014},
   number={16},
   journal={Class. Quantum Gravity},
   publisher={IOP Publishing},
   author = {A. Ashtekar and M. Bojowald},
   year={2005},
   month=aug, pages={3349–3362} }

@article{Ashtekar_2020,
   title={Black Hole Evaporation: A Perspective from Loop Quantum Gravity},
   volume={6},
   ISSN={2218-1997},
   url={http://dx.doi.org/10.3390/universe6020021},
   DOI={10.3390/universe6020021},
   number={2},
   journal={Universe},
   publisher={MDPI AG},
   author = {A. Ashtekar},
   year={2020},
   month=jan, pages={21} }

@article{Ashtekar_2025,
   title={Black hole evaporation in loop quantum gravity},
   volume={57},
   ISSN={1572-9532},
   url={http://dx.doi.org/10.1007/s10714-025-03380-7},
   DOI={10.1007/s10714-025-03380-7},
   pages={48},
   journal={Gen. Relativ. Gravit.},
   publisher={Springer Science and Business Media LLC},
   author = {A. Ashtekar},
   year={2025},
   month=feb }

@misc{strominger2017blackholeinformationrevisited,
      title={Black Hole Information Revisited}, 
      author = {A. Strominger},
      year={2017},
      eprint={1706.07143},
      archivePrefix={arXiv},
      primaryClass={hep-th}}

@article{Chen_2015,
   title={Black hole remnants and the information loss paradox},
   volume={603},
   ISSN={0370-1573},
   url={http://dx.doi.org/10.1016/j.physrep.2015.10.007},
   DOI={10.1016/j.physrep.2015.10.007},
   journal={Physics Reports},
   publisher={Elsevier BV},
   author = {P. Chen and Y. Ong and D. Yeom},
   year={2015},
   month=nov, pages={1–45} }

@article{Mathur_2005,
   title={The fuzzball proposal for black holes: an elementary review},
   volume={53},
   ISSN={1521-3978},
   url={http://dx.doi.org/10.1002/prop.200410203},
   DOI={10.1002/prop.200410203},
   number={7–8},
   journal={Fortschr. Phys.},
   publisher={Wiley},
   author = {S. Mathur},
   year={2005},
   month=jun, pages={793–827} }

@article{petr_kay_kuchar1992quantum,
  title = {Quantum collapse of a self-gravitating shell: Equivalence to Coulomb scattering},
  author = {P. H{\'a}j{\'\i}{\v{c}}ek and B. S. Kay and Kucha{\v{r}}
, K. V.},
  journal = {Phys. Rev. D},
  volume = {46},
  issue = {12},
  pages = {5439--5448},
  numpages = {0},
  year = {1992},
  month = {Dec},
  publisher = {American Physical Society},
  doi = {10.1103/PhysRevD.46.5439},
  url = {https://link.aps.org/doi/10.1103/PhysRevD.46.5439}
}

@article{Vaz_2022,
   title={Quantum collapse of a thin shell revisited},
   volume={105},
   ISSN={2470-0029},
   url={http://dx.doi.org/10.1103/PhysRevD.105.086020},
   DOI={10.1103/physrevd.105.086020},
   pages={086020},
   journal={Phys. Rev.  D},
   publisher={American Physical Society (APS)},
   author = {C. Vaz},
   year={2022} }

\end{document}